\documentclass[11pt]{article}
\usepackage[a4paper,hmarginratio=1:1,vmarginratio=2:3,totalwidth=15.5cm,
totalheight=21.45cm]{geometry}
\usepackage{bm,epstopdf,epsfig,amsmath,amssymb,amsfonts,latexsym,cancel,verbatim,ulem}

\usepackage{epsfig} 
\usepackage[usenames,dvipsnames]{color}
\usepackage{graphicx}
\renewcommand{\baselinestretch}{1.15}

\newcommand{\cL}{{\cal L}}

\newcommand{\cO}{{\cal O}}

\newcommand{\cZ}{{\cal Z}}

\newcommand{\ra}{\rightarrow}
\newcommand{\be}{\begin{equation}}
\newcommand{\ee}{\end{equation}}
\newcommand{\bea}{\begin{eqnarray}}
\newcommand{\eea}{\end{eqnarray}}

\long\def\symbolfootnote[#1]#2{\begingroup
\def\thefootnote{\fnsymbol{footnote}}\footnote[#1]{#2}\endgroup} 
\begin{document}
\begin{flushright}
CERN-PH-TH-2011-044
\end{flushright}

\thispagestyle{empty}

\vspace{1.2cm}

\begin{center}
{\Large {\bf A review of naturalness and dark matter prediction 

\bigskip

for the Higgs mass in MSSM and beyond}}

\vspace{1.cm}
\textbf{S. Cassel$^{\,\,a}$,  
D.~M. Ghilencea$^{\,\,b,\,c}$
\symbolfootnote[2]{E-mail addresses:
 s.cassel1@physics.ox.ac.uk, dumitru.ghilencea@cern.ch}} \\

\vspace{0.7cm} 
{\small $^a\, $ Rudolf Peierls Centre for Theoretical Physics, University of Oxford,\\[0pt]
1 Keble Road, Oxford OX1 3NP, United Kingdom.}\\

\vspace{0.3cm}
{\small $^b\, $ CERN - Theory Division, CH-1211 Geneva 23, Switzerland.}\\

\vspace{0.3cm}
{\small $^c\, $ Theoretical Physics Department, \\[0pt]
National Institute of Physics and Nuclear Engineering (IFIN-HH),
Bucharest MG-6, Romania.}\\

\end{center}

\def\baselinestretch{1.}
\begin{abstract}
Within  a two-loop leading-log approximation, we review  the prediction  for the lightest 
Higgs mass ($m_h$) in the  framework of constrained MSSM (CMSSM), derived from the 
naturalness requirement of minimal fine-tuning ($\Delta$) of the electroweak scale, and 
dark matter consistency.  As a result, the Higgs  mass  is predicted to be just above 
the LEP2 bound, $m_h=115.9\pm 2$ GeV, corresponding to a  minimal $\Delta=17.8$,
value obtained from  consistency with electroweak and WMAP (3$\sigma$) constraints, 
but {\it without} the LEP2 bound. Due  to quantum corrections (largely QCD ones for $m_h$
above LEP2 bound), $\Delta$ grows $\approx$ ~exponentially on either side of  the above 
value of $m_h$, which stresses the relevance  of this prediction. A value
$m_h\!>\!121$ ($126$) GeV cannot be accommodated within the CMSSM  
unless one accepts  a fine-tuning cost worse than $\Delta\!>\!100$ ($1000$), respectively. 
We review how the above prediction for $m_h$ and $\Delta$ changes under the addition of new physics 
beyond the MSSM Higgs sector,  parametrized by effective operators  of dimensions d=5 and d=6.  
For d=5 operators, one can obtain values  $m_h\leq 130$ GeV for $\Delta<10$.
The size of the supersymmetric correction that each individual operator of d=6 
brings  to the  value of $m_h$ for points with $\Delta<\!100$ ($<\!200$), is found   to be small, of few  
$\leq 4$ GeV ($\leq 6$ GeV) respectively, for $M=8$ TeV where $M$ is the scale of new physics.  
This value  decreases (increases) by approximately 1 GeV for a 1 TeV increase 
(decrease) of the scale~$M$. The relation of these results 
to the Atlas/CMS supersymmetry exclusion limits  is presented 
together with  their impact for the CMSSM regions of  lowest fine-tuning. 
\end{abstract}

\newpage

\section{Naturalness measures and predictions.}

After forty years of supersymmetry, we  are  fortunate to currently witness  its 
biggest real test at the Large Hadron Collider (LHC). It remains to be seen if the LHC
experiments will prove the idea of low-energy (TeV-scale) supersymmetry, or 
just increase its scale to higher values.
Negative searches for superpartners in the TeV region can ultimately question it 
as a natural solution to the hierarchy problem. This solution resides on quantum  
cancellations between partners and superpartners, that becomes increasingly difficult (in 
the absence of some ``tuning'' 
 of the soft scales) when the latter are too far above the TeV scale. 
In that case the stability of the electroweak (EW) scale under these quantum effects,
with well above TeV-scale spartners may become questionable.  A way to see 
this more quantitatively 
is to examine the stability (or relative variation)  of the EW scale under (relative) 
variations of the UV parameters (masses) of the theory,  with all experimental 
constraints taken into account.
This relative variation is the so-called EW fine-tuning measure introduced long 
ago \cite{Ellis:1986yg} and is a quantitative measure 
of supersymmetry  as a solution to the hierarchy problem.

A highly stable EW scale under the aforementioned quantum corrections is indeed
preferable, which indicates that the fine tuning measure should be minimal 
with respect to (variation of) 
these UV parameters and this fact will enable us to make {\it predictions}. 
The corresponding region of the parameter space can then be regarded as the most natural, 
and its phenomenological predictions are worth a careful investigation. 
Adding to this analysis the requirement that the results found
satisfy the  WMAP constraint \cite{wmap} will enable us
to put  together large  distance (dark matter) and short distance (EW/TeV scale) 
physics, and thus to improve the consistency and the predictive power of such study.

The results of such an investigation, at two-loop leading-log level, are reviewed below, 
in the context of constrained MSSM. As we shall see later, the need for a two-loop
 level calculation instead of a 1-loop one is well motivated. This, together with a
 careful account of the dependence of $\Delta$ on the MSSM parameters, will bring 
interesting, new results. We then go beyond the MSSM framework, by considering
the effects on the EW scale fine-tuning and Higgs mass 
from ``new physics'' that may exist beyond this model
and which we parametrize using effective operators.
For technical  details of the results we present
see \cite{Cassel:2010px,Antoniadis:2009rn}.

One can argue that we are  missing a quantitative indicator of
 what upper level of fine-tuning is still acceptable
 or of  what makes a model more natural than another. 
We shall not address the latter issue, but for a given model one can accept that,
while a given value of $\Delta$ may be a subjective criterion of naturalness, the best 
$\Delta$ is certainly the one that is minimized with respect to the UV 
parameters of the model, as 
explained above. This is the point of view that we adopt and this  has some support from
the Bayesian method to data fitting \cite{AbdusSalam:2009qd} in which $\Delta$, as 
defined in \cite{Ellis:1986yg} and used below, is  automatically 
present~\cite{Allanach:2006jc,CS}.  
Indeed, in this method $1/\Delta$   emerges naturally
as an {\it effective} prior \cite{CS} in the analysis 
of a  precisely measured  observable, in this case $m_Z$ (the EW scale). 
In this way the prior and therefore the Bayesian method impose automatically
a fine-tuning ``penalty'' for those points of larger $\Delta$, which are in 
this way excluded from the final fit.  This shows the physical 
meaning of $\Delta$ and the Bayesian preference for points 
of small $\Delta$. We shall then search for points of minimal $\Delta$.

To place this discussion on more quantitative grounds
 consider the MSSM scalar  potential
\begin{eqnarray} 
V&=&  m_1^2\,\,\vert H_1\vert^2
+  m_2^2\,\,\vert H_2\vert^2
- (m_3^2\,\,H_1 \cdot H_2+h.c.)\nonumber\\[3pt]
 &&
 ~+~
(\lambda_1/2) \,\vert H_1\vert^4
+(\lambda_2/2) \,\vert H_2\vert^4
+\lambda_3 \,\vert H_1\vert^2\,\vert H_2\vert^2\,
+\lambda_4\,\vert H_1\cdot H_2 \vert^2\nonumber\\[3pt]
 &&
 ~+~
\Big[\,(\lambda_5/2)\,\,(H_1\cdot  H_2)^2
+\lambda_6\,\,\vert H_1\vert^2\, (H_1 \cdot H_2)+
\lambda_7\,\,\vert H_2 \vert^2\,(H_1 \cdot H_2)+h.c.\Big]
\label{2hdm}
\end{eqnarray}

\medskip\noindent
The couplings $\lambda_j$ and the soft masses receive
one- and two-loop corrections that for the MSSM are found
in \cite{Martin:1993zk,Carena:1995bx}. 
The fine tuning of the EW scale with respect to a set of parameters $p$ 
is~\cite{Ellis:1986yg} 
\begin{equation} 
\Delta \equiv \max \big\vert \Delta _{p}\big\vert_{p=\{\mu 
_{0}^{2},m_{0}^{2},A_{0}^{2},B_{0}^{2},m_{1/2}^{2}\}},\qquad \Delta _{p}\equiv \frac{\partial \ln
  v^{2}}{\partial \ln p}  \label{ft} 
\end{equation}

\medskip\noindent
in a standard notation for the constrained MSSM parameters $p$.

As mentioned earlier, one could ask whether this otherwise widely used formula
 for fine-tuning is the most appropriate.  Variations of this definition indeed exist: 
for example another possibility is to use the ``quadrature'' version of $\Delta$ 
defined as
\bea\label{ss34}
\Delta^\prime=\Big\{\,\, {\sum_{p} \Delta_p^2}\,\,\Big\}^{1/2}
\eea
It  is generally agreed that within a model, a value of $\Delta$ or $\Delta^\prime>\!100$ 
or so  (i.e. fine tuning
worse than  1 part in 100) is rather unacceptable. In any case, one would prefer
 to have a minimization 
of $\Delta$ or $\Delta^\prime$ with respect to the above UV parameters $p$.
 Using this idea,   
we  identify  the regions of the CMSSM parameter space of minimal $\Delta$ that 
can then be regarded as the  most natural. 
This idea can be used to make {\it predictions}, such as to find the most natural 
value of the SM-like Higgs mass $m_h$, which is important. 
The results given in the following are all based on definition (\ref{ft}),
however we checked  that the predicted value of $m_h$ obtained from  minimizing 
instead  $\Delta^\prime$ is the same, which we find interesting.

Let us present the general idea briefly, developed further 
in the next section. In the CMSSM, at tree level $m_h\!\leq\! m_Z$, 
 and even in the presence  of quantum corrections, 
 $m_h$  is often  below the  observed LEP2 bound \cite{LEP2}
($114.4$ GeV), however large values for $m_h$ (up to $\approx 135$ GeV) can indeed be achieved.  
One would like to clarify what  value for $m_h$  in this range ($m_Z, 135$ GeV) 
is the most natural, in the light of the criterion of minimal $\Delta$ mentioned and 
in the {\it absence} of the LEP2 bound, that  is actually  not imposed in the following. 
By doing so, we explore the whole range of quantum corrections to $m_h$, from 
their vanishing to their largest values, without the prejudice of
imposing a cut on their size, as LEP2 bound would  do. 
We return to this problem later in the text, when we discuss the dependence of 
$\Delta$ on radiatively corrected $m_h$ and the minimum of $\Delta$ for the whole 
parameter space.

One of the two minimum conditions of the scalar potential $V$ in MSSM gives 
$v^2=-m^2/\lambda$ where
$m^{2}=m^{2}(p,\beta )$ is a combination  of soft masses ($m_{1,2,3}$) while
$\lambda =\lambda (p,\beta )$ is the {\it effective} quartic Higgs coupling:
\medskip
\begin{eqnarray}
m^2 &=&
 m_1^2 \, \cos^2 \beta +  m_2^2
 \, \sin^2 \beta - m_3^2 \, \sin 2\beta\nonumber\\[3pt]
\lambda &=&\frac{ \lambda_1^{} }{2} \, \cos^4 \beta 
+ \frac{ \lambda_2^{} }{2} \,  \sin^4 \beta 
+ \frac{ \lambda_{345}^{} }{4} \, \sin^2 2\beta 
+ \sin 2\beta \left( \lambda_6^{} \cos^2 \beta 
+ \lambda_7^{} \sin^2 \beta \right)
\label{ml}
\end{eqnarray} 

\medskip\noindent
where $\lambda_{345}=\lambda_3+\lambda_4+\lambda_5$.
Let us examine the first minimum condition for $V$: $v^2=-m^2/\lambda$. 
The  problem one sees immediately is that
with $v\sim \cO(100)$ GeV, $m\sim \cO(TeV)$ but with  $\lambda\leq 1$ it is in 
general difficult to  satisfy this minimum condition.
 To this purpose, with $v$ fixed to the EW scale,
one has  to keep $m^2$ as low as possible, closer to EW (scale)$^2$, by 
``tuning'' the loop-corrected coefficients entering in the first eq in (\ref{ml})
so that the near-TeV scales present there largely cancel together, 
to leave $m\sim \cO(100)$  GeV. 
Another way to phrase this  problem (sometimes called ``little hierarchy'') 
is how to separate the EW and supersymmetry breaking scale 
($m^2$), while still respecting the minimum condition
$v^2=-m^2/\lambda$, with TeV-scale soft masses.
This ``tension'' is only one aspect captured by the fine-tuning measure $\Delta$, and is  
usually more emphasized  compared to a more important one, 
related to the size of $\lambda$, discussed next.

Obviously, any increase of $\lambda$ will reduce the aforementioned tension and ultimately
will reduce  $\Delta$. This increase can be  due to quantum corrections to the couplings $
\lambda_i$,  or to other corrections to them 
from ``new physics''.  This stresses the importance of quantum corrections to $\lambda$
in the MSSM. As for the ``new physics'' beyond the MSSM, this can be represented by new gauge interactions,
 additional effective couplings due to integrating out some massive states, etc. 
Note that the smallness of $\lambda$, fixed at the tree level by SM gauge interactions, 
is partly to blame  for this fine-tuning problem, in addition to the larger values of
 soft masses ($m_{1,2,3}$), from negative Susy searches or other effects 
(like demanding the largest quantum  correction to the Higgs mass to respect the LEP2 bound). 
Thus the (``little hierarchy'') problem discussed here is not only one of
 mass scales (EW versus Susy breaking scale) but also of the smallness of 
couplings $\lambda_i$.

We can summarize this discussion by remarking
that there  are two competing effects, quantum corrections or ``new physics''
contributions that can increase $m^2$ (or $m^2_{1,2,3}$) against those that could increase
effective  $\lambda$.  It is indeed possible that the latter corrections dominate in some 
region of parameter space, to allow low fine-tuning, and we shall identify 
this region. Ultimately, if one keeps increasing
the soft masses to very large values, these will dominate the fine-tuning, 
to the extreme case of recovering the SM case and the hierarchy problem.

There is a second minimum condition of $V$ that can be written in the form
\medskip
\bea
2\lambda \frac{\partial m^2}{\partial \beta}=m^2\frac{\partial\lambda}{\partial\beta}
\eea

\medskip\noindent
This condition induces an implicit dependence 
of $\tan\beta$ on parameters $p$ (via $\lambda=\lambda(p,\beta)$),
that must be carefully taken into account when evaluating $\Delta$ 
\cite{Cassel:2010px};  one can then trade $B_0$ for $\tan\beta$ which can
then be regarded as an independent parameter.
With this information, one can show that the fine tuning has a general formula
\cite{Casas:2003jx} (see also the Appendix of \cite{Cassel:2009ps}): 
\medskip
\begin{eqnarray} \label{delta0}
\Delta _{p} &=& -\frac{p}{z}\,\bigg[\bigg(2
\frac{\partial ^{2}m^{2}}{\partial 
\beta ^{2}}+v^{2}\frac{\partial ^{2}\lambda }{\partial \beta ^{2}}\bigg)
\bigg(\frac{\partial \lambda }{\partial p}+\frac{1}{v^{2}}\frac{\partial 
m^{2}}{\partial p}\bigg)+\frac{\partial m^{2}}{\partial \beta }\frac{
\partial ^{2}\lambda }{\partial \beta \partial p}-\frac{\partial \lambda }{
\partial \beta }\frac{\partial ^{2}m^{2}}{\partial \beta \partial p}\bigg]
\nonumber
\\[6pt]
z &=& \lambda \,\bigg(2\frac{\partial ^{2}m^{2}}
{\partial \beta ^{2}}+v^{2}\frac{ 
\partial ^{2}\lambda }{\partial \beta ^{2}}\bigg)-\frac{v^{2}}{2}\,\bigg(
\frac{\partial \lambda }{\partial \beta }\bigg)^{2}
\end{eqnarray} 

\bigskip\noindent
This is the most general dependence of $\Delta$ on the parameters $p$ of the model
and has the advantage that once the radiative corrections to the 
couplings and masses are introduced, can easily be evaluated.
Using available 2-loop leading log corrections for couplings entering the 
MSSM Higgs potential \cite{Martin:1993zk,Carena:1995bx} one ends up with 
a general expression for fine tuning which can be used to evaluate 
$\Delta$ at two-loop. For technical details on this matter see \cite{Cassel:2010px}.

We would like to stress the strong effects that quantum corrections to couplings 
$\lambda_i$ have in reducing the fine tuning. Note that in the limit these corrections 
to $\lambda_i$ are turned off, one can show \cite{Cassel:2010px} that after some 
calculations $\Delta$  of (\ref{delta0}) reduces to the so-called ``master formula'' 
 of \cite{Dimopoulos:1995mi} (see also 
\cite{Chankowski:1997zh,Chankowski:1998xv}), which  is
\bigskip
\bea
\Delta_{p}\!& =& \!
\frac{-p\cos^2\beta}{m_Z^2\cos2\beta}
\bigg\{\frac{\partial  m_1^2}{\partial p}\!-\!
\tan^2\beta\frac{\partial  m_2^2}{\partial p}
\nonumber\\[7pt]
&&\qquad\qquad
- \frac{\tan\beta}{\cos 2\beta}\bigg[1
\!+\!\frac{m_Z^2}{m_1^2\!+\! m_2^2}\bigg]
\bigg[2\frac{\partial m_3^2}{\partial p}\!-\!\sin 2\beta
\Big(\frac{\partial m_1^2}{\partial p}\!+\!
\frac{\partial m_2^2}{\partial p}\Big)
\bigg]\bigg\}\label{gd}\!\!\! \qquad
\eea

\bigskip\noindent
The numerical discrepancy between the result given by (\ref{gd}) and the 
more general formula (\ref{delta0})  is indeed 
significant and  (\ref{gd}) often gives larger $\Delta$ than (\ref{delta0}).
Unfortunately, most works in the literature use (\ref{gd}), albeit with loop corrections to
the effective couplings later included in the potential, but not in $\Delta$; 
this lead to significant overestimates of the overall fine-tuning amount. 
This problem can be avoided if using SoftSusy \cite{Allanach:2001kg} 
 when the whole procedure can be done numerically and agrees well with
 (\ref{delta0}) \cite{Cassel:2010px} in the 2-loop leading log approximation.

To illustrate the discrepancy between the two expressions, 
it can be shown that including only the one-loop correction $\delta$
from stop/top Yukawa  coupling to our Higgs coupling 
$\lambda_2\ra \lambda_2(1+\delta)$ and then using (\ref{delta0}), one finds
\bigskip\bea
\Delta_p\propto \frac{p}{(1+\delta)\,m_Z^2+\cO(1/\tan\beta)}, 
\qquad p=\mu_0^2,
m_0^2, m_{1/2}^2, A_0^2, B_0^2.
\label{fggf}
\eea

\bigskip\noindent
Since usually $\delta=\cO(1)$, one sees that a factor of 2 reduction 
in $\Delta$ is easily achieved by $\delta$ (effect not captured by (\ref{gd})) 
even in this limiting case of only one Yukawa correction. 
More corrections to the couplings $\lambda_i$ are  likely to reduce $\Delta$ further.
Such effects were not included in the previous estimates of fine-tuning, and are likely to
reduce the CMSSM overall amount of fine-tuning, as we shall see shortly.

There is another way to illustrate the strong impact of quantum corrections on $\Delta$.  
Usually $\Delta\sim m_0^2\sim \exp(m_h^2/m_Z^2)$, where the first step comes from first 
eq in (\ref{ml}), while the second from the fact that leading quantum corrections to $m_h$ 
are of the type $m_h^2\sim \log m_0^2$, which is then inverted into an exponential. 
Therefore  $\Delta$ depends exponentially on the quantum corrections to $m_h$.
 Even though 2-loop leading-log corrections to $m_h$ may be small 
relative to the 1-loop ones, when ``exponentiated'' they can bring a significant impact 
on $\Delta$. For this reason, it is advisable to include not only 1-loop but also 2-loop 
corrections to $\lambda_i$.

In the light of the exponential behaviour of $\Delta$ wrt $m_h$, 
the very existence of a global minimum of $\Delta$ situated at the 
intersection of two such exponential 
dependences on $m_h$ (see later) that follow the
 two minimum conditions, cannot be stressed enough. 
The results reviewed below  always use the general formula eq.(\ref{delta0}) 
with 2-loop  leading-log corrections to~$\lambda_i$.

\section{$m_h$ from minimal fine-tuning and dark matter consistency.}\label{section3}

With the above observations, we review the implications of minimal $\Delta$
for the whole  parameter space. Associating this with the most natural
 regions of parameters values, we can see what this criterion predicts for the Higgs sector.
The numerical results  for $\Delta$ include two-loop corrections with
the theoretical constraints:  radiative  electroweak breaking (EWSB), non-tachyonic
sparticles masses (avoiding colour and charge breaking (CCB) vacua),
and experimental constraints: bounds on superpartner
masses, electroweak precision data, $b\ra s\,\gamma$, $B_s\ra \mu^+\,\mu^-$
and anomalous magnetic moment $\delta a_\mu$. For further details and for 
the actual experimental values considered see  Table~1 in \cite{Cassel:2010px}. 
Let us stress that, unless stated otherwise, 
consistency of $m_h$ with the LEP2 bound (114.4 GeV \cite{LEP2}) and/or 
consistency with the dark matter abundance (of the LSP), are {\it not} imposed. 
Later we discuss separately their impact on the value of predicted $m_h$.

\begin{figure}[t] 
\begin{center}
\begin{tabular}{cc|cr|} 
\parbox{6.5cm}{\psfig{figure=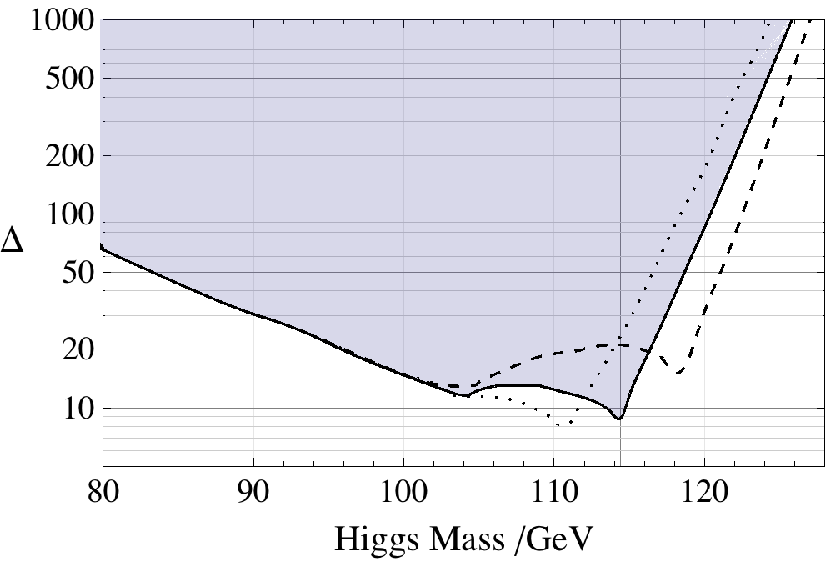,
height=5.cm,width=6.4cm}} \hspace{0.4cm}
\parbox{6.5cm}{\psfig{figure=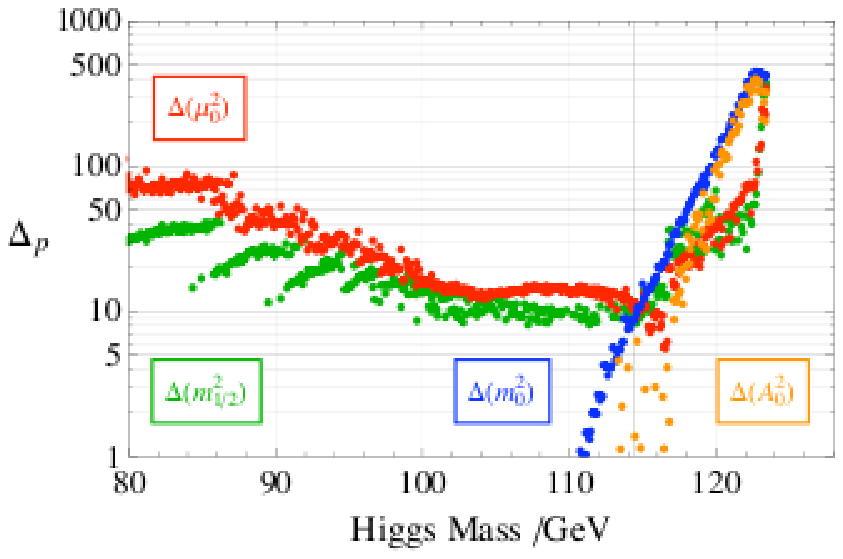,
height=5.cm,width=6.4cm}}
\end{tabular}
\end{center}
\renewcommand{\baselinestretch}{1.}
\caption{\small
Left figure: Fine tuning vs Higgs mass, in a two-loop analysis, 
for a wide range of parameters $\mu_0, m_0, A_0, B_0, m_{1/2}$ 
and for $2 \leq \tan \beta \leq 55$.
The solid line is the minimum fine tuning with central values 
for strong coupling ($\alpha_3(m_Z)$) and top mass ($m_t$):
$(\alpha_3, m_t)=(0.1176,173.1$\,GeV).  The dashed line corresponds to
 $(\alpha_3, m_t)=(0.1156,174.4$\,GeV) and the dotted line to
$(0.1196,171.8$\,GeV), to account for   $1\sigma$
variations of $\alpha_3$ and top mass \cite{TeV}.
The LEP2 bound of $114.4$~GeV is indicated by a vertical line. 
Note the steep ($\approx$ exponential) increase of $\Delta$ 
on both sides of the minimum value, situated near the LEP2 bound for $m_h$.
Right figure: The plot displays, for any fixed $p$,  
$\max\vert\Delta_p\vert$, $p=\mu_0^2,m_0^2, A_0^2, B_0^2, m_{1/2}^2$, that contribute
to overall $\Delta$ of  the left figure.  The largest of these all, for all $p$ and 
$m_h$ gives the boundary contour presented  in the left figure.}\label{dplot}
\end{figure}

The results on fine tuning presented below were obtained using a combined analytical 
and numerical approach, using 2-loop leading log corrections implemented as in 
\cite{Cassel:2010px}. The results obtained were in very good agreement with those found 
using SoftSusy code, whose very long CPU run time was reduced (from 6-years on 
30$\times$3GHz), using a specially designed  Mathematica code. 
This was one reason that prevented earlier investigations of $\Delta$ of similar accuracy.
The Mathematica code was used to select the phase space points that 
respected the above constraints and also to identify points of minimal $\Delta$, 
which were then used in SoftSusy and this  reduced the run time to manageable levels.

The results obtained are plotted in  Figure~\ref{dplot}.
In the left figure, 2-loop $\Delta$ is presented as a function of the Higgs mass.  
Interestingly, there is a minimum of $\Delta\approx 8.8$ 
which predicts a value of $m_h$ which is just above the LEP2 bound, at $m_h=114\pm 2$ GeV.
The quoted theoretical uncertainty of $\pm 2$ GeV is due to  higher 
order perturbative corrections that account for differences between 
the results of SoftSusy  and FeynHiggs codes and can be even larger, up to 3 GeV.
This value of $m_h$ is found with  the above experimental 
and theoretical constraints, but  without imposing the LEP2 bound. It just 
turns out that minimal fine-tuning wrt to UV parameters indicates a 
size for the quantum corrections that prefer a total value for $m_h$ close 
to this bound.

 Notice the presence in  Figure~\ref{dplot} of the steep, $\approx$ exponential 
increase of minimal $\Delta$  on both sides of  its minimum value situated near the LEP2 
bound,  effect largely due to quantum corrections (note the log OY scale). 
This behaviour underlines the importance  of the minimal value of $\Delta$.
Variations of $1\sigma$ around the central values of $\alpha_3(m_Z)$ 
and of the top mass, give the results shown by the dotted and dashed lines: 
a 1$\sigma$ increase of $\alpha_3(m_Z)$ gives very similar results to a 1 $\sigma$ 
decrease of top mass; thus these have opposite effects. Indeed a larger top Yukawa helps 
a radiative  EWSB, while a larger strong coupling $\alpha_3(m _Z)$ has an opposite effect, at two-loop. 
The minimum of $\Delta$ is where these effects are balanced, in other words - assuming 
minimal $\Delta$, the Higgs mass prediction cannot be larger due to QCD quantum effects.
In other words, QCD does not ``like'' a larger Higgs mass, unless one is prepared 
to accept the fine tuning cost that comes with it (this is indeed very high for $m_h> 126$ GeV, 
when one already has $\Delta>1000$). 
The conclusion is that, from a 2-loop evaluation of $\Delta$, one finds, rather 
intriguingly,  that minimal $\Delta\approx 8.8$ favours the value 
$m_h=114\pm 2$ GeV.

\begin{figure}[t!] 
\begin{center}
\begin{tabular}{cc|cr|} 
\parbox{6.5cm}{\psfig{figure=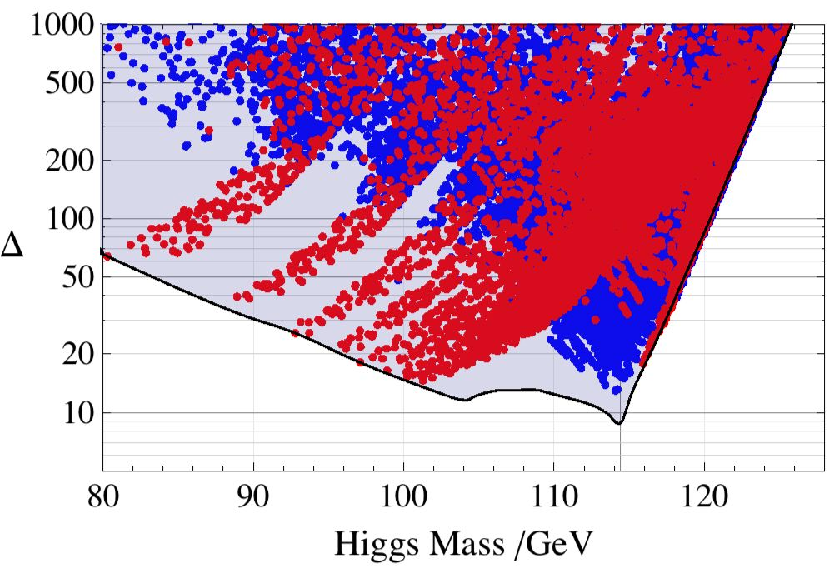,
height=5cm,width=6.4cm}} \hspace{0.4cm}
\parbox{6.5cm}{\psfig{figure=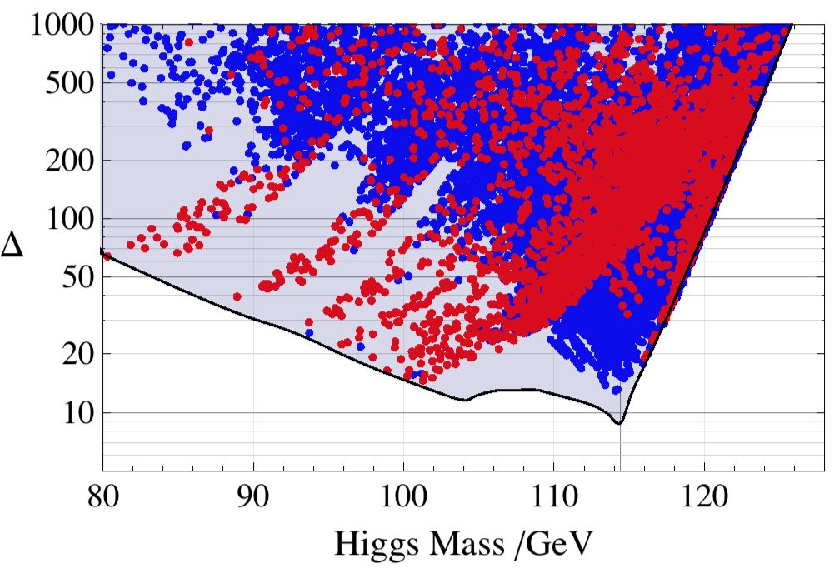,
height=4.7cm,width=6.4cm}}
\end{tabular}
\end{center}
\renewcommand{\baselinestretch}{1.}
 \caption{\small 
Left plot: Two-loop fine tuning vs Higgs mass with the influence of the WMAP bound. 
The blue (darker) points sub-saturate the relic density.
 The red (lighter) points correspond to a relic density within the
3$\sigma$ bounds of $\Omega h^2 = 0.1099\pm0.0062$ \cite{wmap}. The `strips' of
points at low Higgs mass appear due to taking steps of 0.5 in $\tan
\beta$ below 10. A denser scan is expected to fill in this
region. Similarly, more relic density saturating points are expected
to cover the wedge of sub-saturating points at $m_h^{}\sim114$~GeV and 
$\Delta \gtrsim 30$. 
The continuous line is that of minimal electroweak $\Delta$ of  Fig.~\ref{dplot},
 without the relic density constraint.
Right plot: as for left plot, but within 1$\sigma$. 
}
 \label{omcon}
 \end{figure}

\begin{figure}[tbh] 
\begin{center}
\begin{tabular}{cc|cr|} 
\parbox{6.9cm}{\psfig{figure=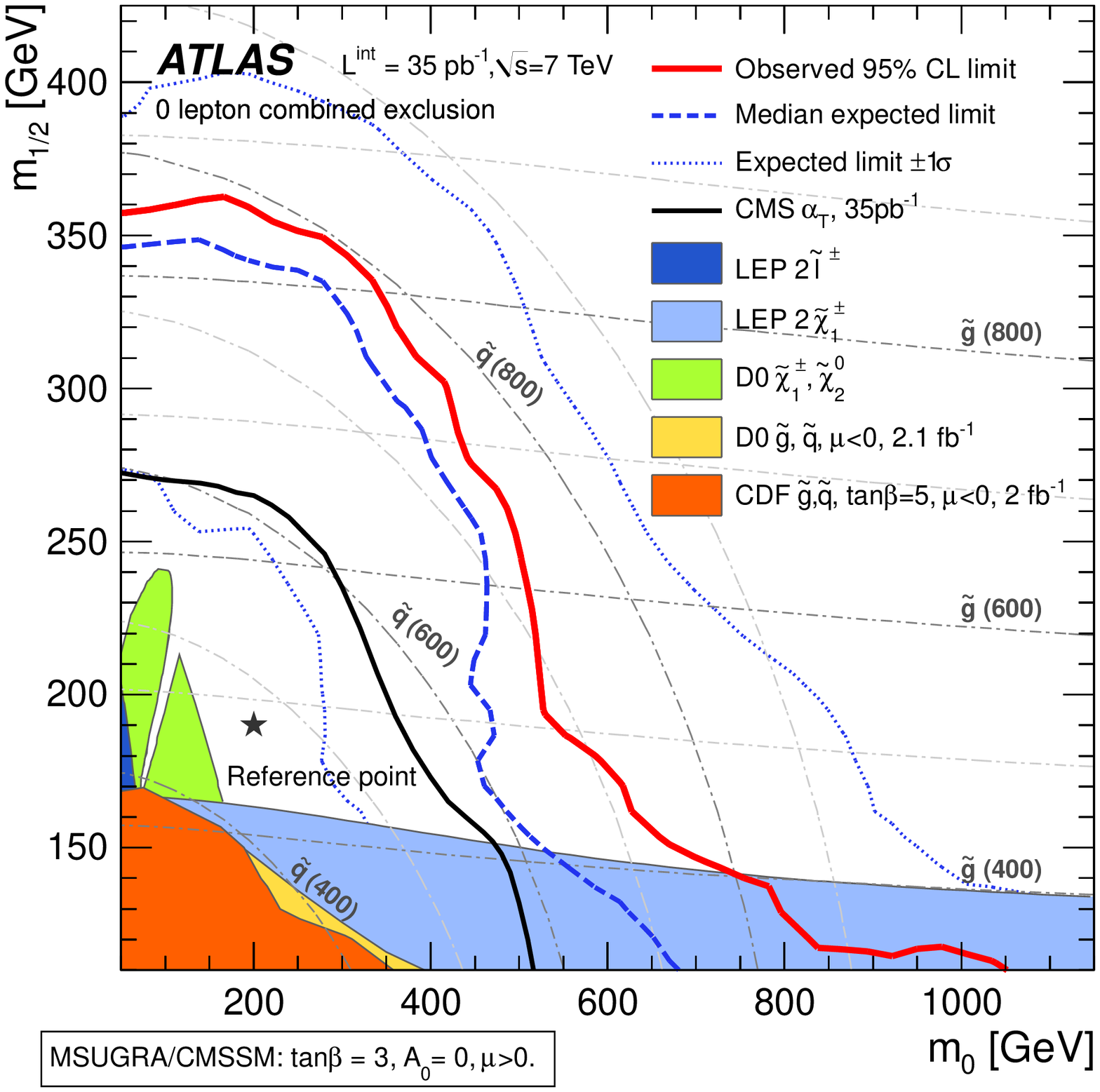,
height=5.5cm,width=6.9cm}} \hspace{0.4cm}
\parbox{6.9cm}{\psfig{figure=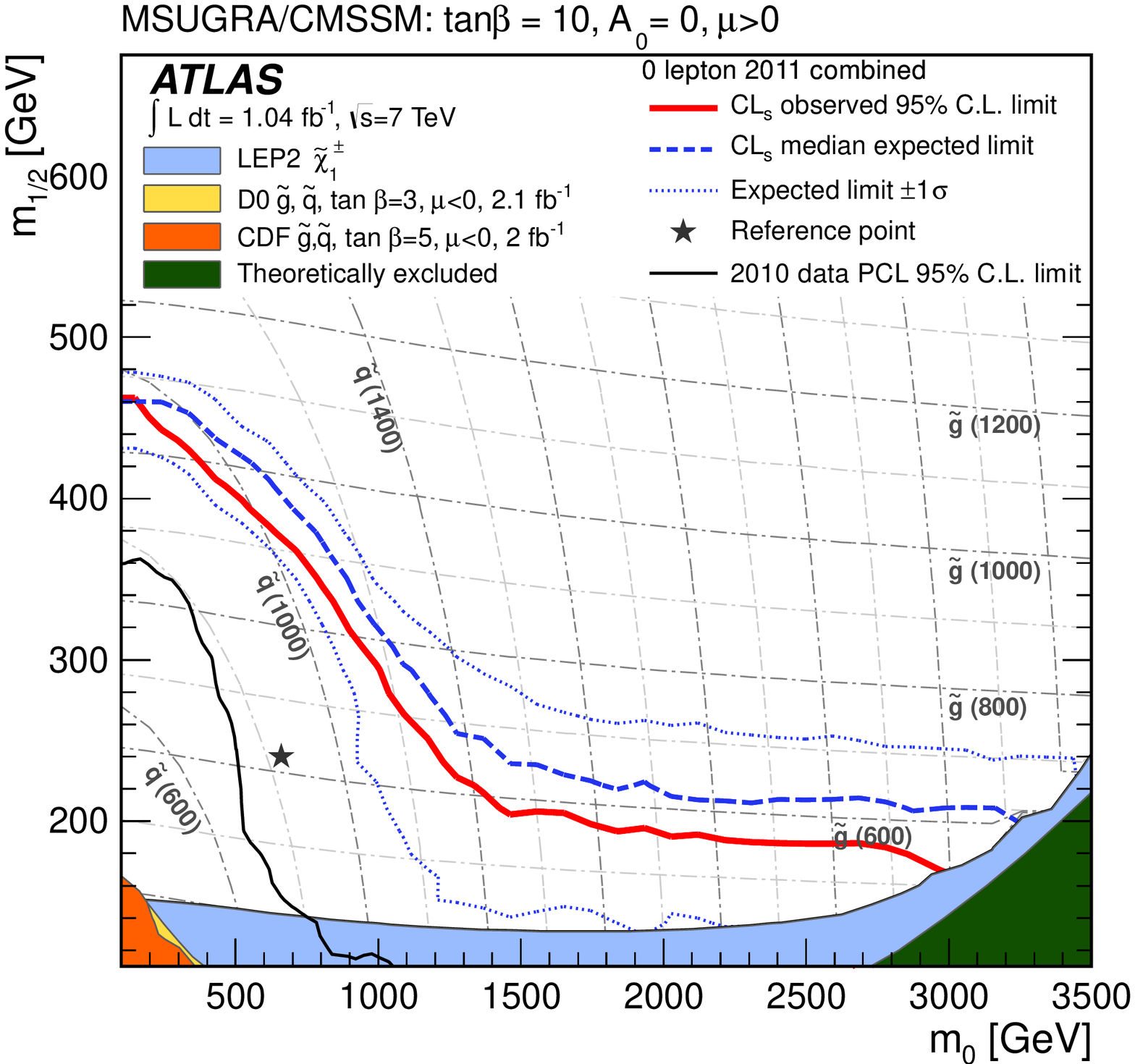,
height=5.5cm,width=6.9cm}}
\end{tabular}
\end{center}
\renewcommand{\baselinestretch}{1.}
 \caption{\small 
Left: The 2010 Atlas (CMS) observed exclusion limit given by the red (back) curve
in the  $(m_0,m_{1/2})$ plane, for $\tan\beta=3$,  $A_0=0$ and $\mu>0$
\cite{daCosta:2011qk}. In our plots of fine tuning in the plane $(m_0,m_{1/2})$,
shown in figures~\ref{omcon2},\ref{oplots},\ref{oq2},
the corresponding red and black exclusion curves are also displayed in similar
colours. Right:
The 2011 Atlas observed exclusion limit (red curve) \cite{arXiv:1109.6572}.
In our plots  of fine tuning in the plane $(m_0,m_{1/2})$ shown in 
Figures~\ref{omcon2}, \ref{oplots}, \ref{oq2}, this 2011
exclusion curve is displayed in green, 
to avoid confusion with the  exclusion curve from the left plot.
Note these exclusion curves are for fixed values of some CMSSM parameters 
($\tan\beta, A_0$, etc).}
 \label{omcon2atlas}
 \end{figure}

\begin{figure}[tbh] 
\begin{center}
\parbox{6.9cm}{\psfig{figure=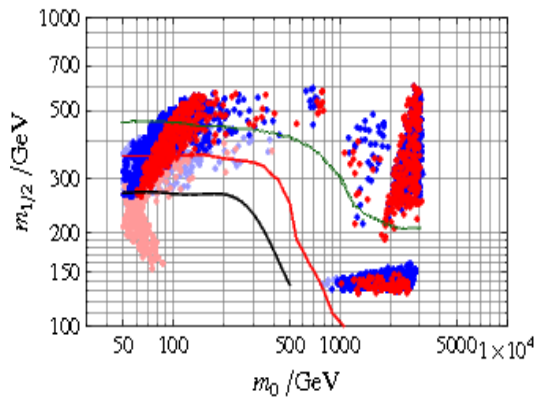,
height=5.5cm,width=6.9cm}}
\end{center}
\renewcommand{\baselinestretch}{1.}
 \caption{\small 
The points in the plane $(m_0,m_{1/2})$ that have $\Delta<100$ 
and are consistent (blue) with the WMAP constraint 
(3$\sigma$ deviation) or saturate it (red) within 3$\sigma$.
These points are the same as in the left plot of Fig.\ref{omcon}.
The points in lighter (darker) red/blue have $m_h$ below (above) the LEP2 bound for $m_h$.
The black, red and green curves correspond to exclusion curves from 
CMS, Atlas 2010 curve \cite{daCosta:2011qk} and Atlas 2011 curve
\cite{arXiv:1109.6572} respectively, see figure~\ref{omcon2atlas} for details.
One sees  that experimental data already test and rule out some
points of low fine tuning $\Delta<100$.}
 \label{omcon2}
 \end{figure}

\begin{figure}[tbh!] 
\begin{center}
\begin{tabular}{cc|cr|} 
\parbox{6.9cm}{\psfig{figure=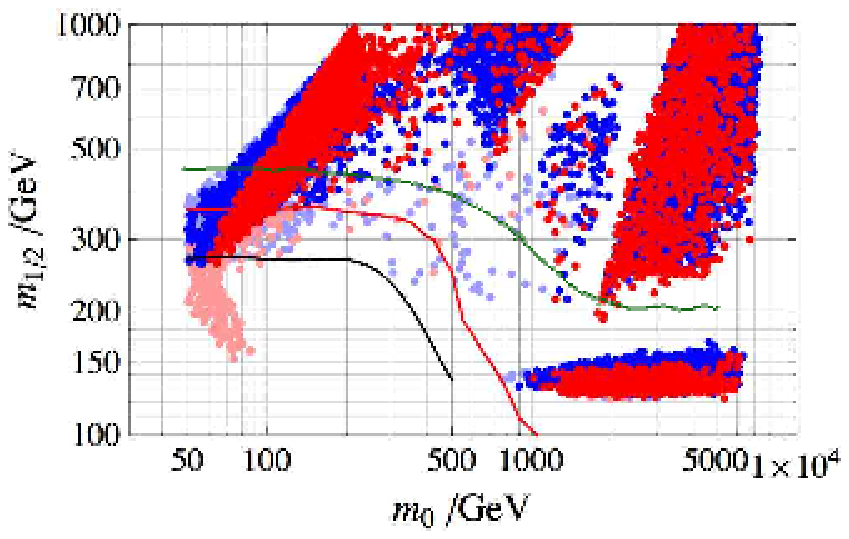,
height=5.5cm,width=6.9cm}} \hspace{0.4cm}
\parbox{6.5cm}{\psfig{figure=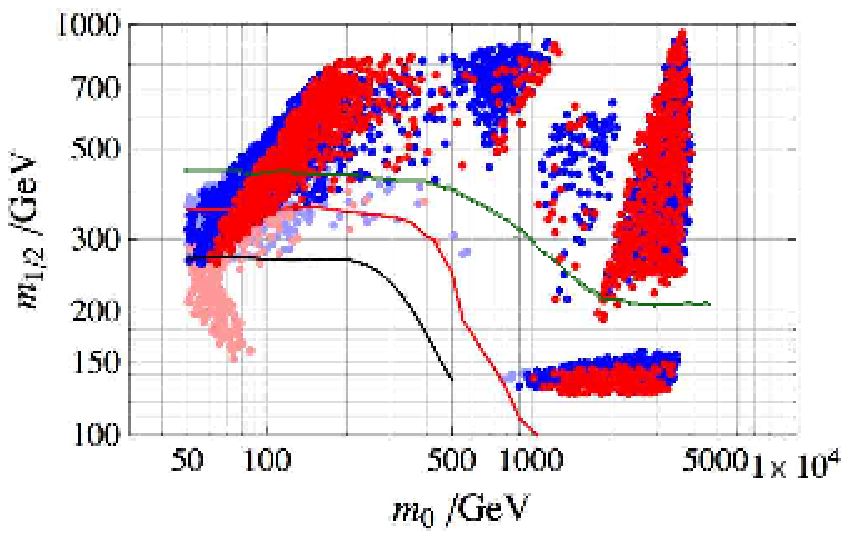,
height=5.5cm,width=6.9cm}}
\end{tabular}
\medskip
\begin{tabular}{cc|cr|} 
\parbox{6.9cm}{\psfig{figure=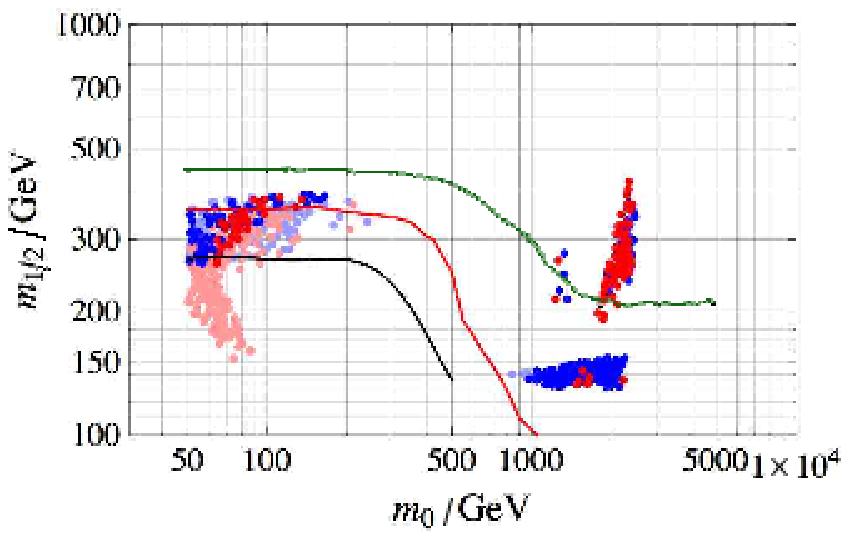,
height=5.5cm,width=6.9cm}} \hspace{0.4cm}
\parbox{6.5cm}{\psfig{figure=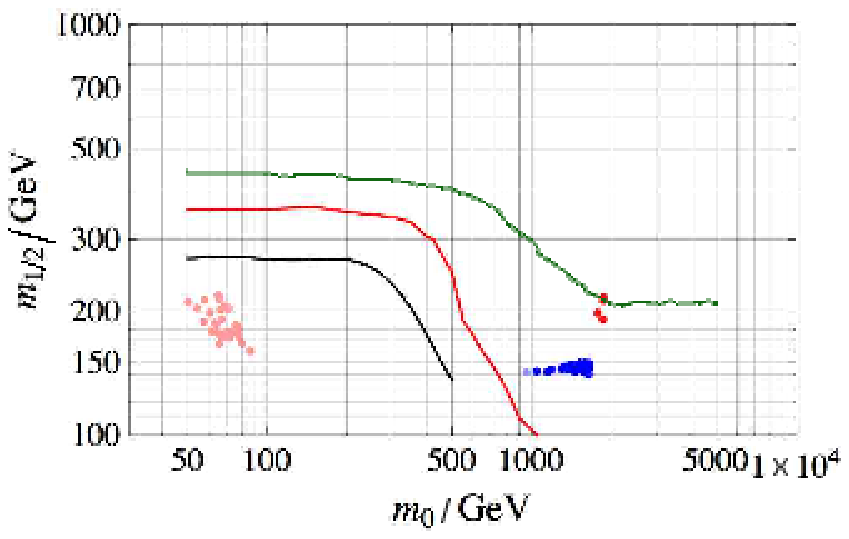,
height=5.5cm,width=6.9cm}}
\end{tabular}
\end{center}
\renewcommand{\baselinestretch}{1.}
 \caption{\small As for the plot in Figure~\ref{omcon2} but with: $\Delta<1000$  
(top left),  $\Delta<200$ (top right),  $\Delta<50$  (bottom 
left plot) and $\Delta<20$ (bottom right plot).
Note that $\Delta<1000$ implies $m_h<126$ GeV, see 
Figure~\ref{dplot}. For $\Delta<10000$, the plot is very similar to that for $\Delta<1000$
with the only difference that the areas near $m_0= 5$ TeV  and any $m_{1/2}$ 
extend up to 10 TeV. See also the plot in Fig.~\ref{omcon2} for $\Delta<100$.
 The points in dark red/blue satisfy the LEP2 lower bound for $m_h$, 
while those in light red/blue do not. This bound is imposed at 111 GeV to
account for the theoretical error of 2-3 GeV mentioned earlier, see text.
The black, red and green curves correspond to exclusion curves from 
CMS, Atlas 2010 curve \cite{daCosta:2011qk} and Atlas 2011 curve
\cite{arXiv:1109.6572} respectively, see figure~\ref{omcon2atlas} for details.}
 \label{oplots}
 \end{figure}

In the right plot of Figure~\ref{dplot} are shown
the individual contributions $\max\vert\Delta_p\vert$,
$p=\mu_0^2$,$m_0^2$, $A_0^2$, $B_0^2$, $m_{1/2}^2$, to the electroweak fine-tuning
$\Delta$. At low $m_h$, below the LEP2 bound $\Delta_{\mu_0^2}$  is dominant, 
while above this bound and at large
$m_h$, $\Delta_{m_0^2}$  is dominant, with $\Delta_{A_0^2}$
reaching similar values near 120 GeV; this contradicts  common 
claims in the literature that  $\Delta_{\mu_0^2}$ is actually the 
largest and dominant part of $\Delta$, (true only below the LEP2 bound).  
The transition between the two regions is happening at about 114.5 GeV, 
shown also in the left plot.  
With $\Delta_{\mu_0^2}$ largely related to the EW effects while $\Delta_{m_0^2}$
 related to QCD  effects, one sees again the significance of 
the minimal value of overall  $\Delta$,
at the interplay of these effects.
  Again, the  LEP2 bound is not imposed at any time.

In principle $\Delta_{h_t^2}$ ($h_t$ is the top Yukawa coupling) 
could  be included in the
overall definition of $\Delta$ of eq.(\ref{ft}). However, its 
contribution to $\Delta$ is always sub-dominant (relative to $\Delta_{m_0^2}$) 
when assuming the  modified definition of $\Delta_p$ \cite{Ciafaloni:1996zh},
appropriate for  measured parameters. Under this modified
definition one must replace:
$\Delta_p\rightarrow \Delta_p\times \sigma_p/p$, where
$\sigma_p$ is the 1$\sigma$ error of experimentally measured $p$, in this case $h_t$.
With this modified definition  $\Delta_{h_t^2}$ does not change $\Delta$. 
See also figure 2
in \cite{Cassel:2010px} where $\Delta_{h_t^2}$ is actually computed and
shown but {\it without} the  modified definition.
The largest $\Delta_p$ for all $m_h$ in the right plot of Figure \ref{dplot}
generates the lower continuous curve of $\Delta$ in the left plot in Fig~\ref{dplot}.

Let us now put together the short distance (EW/TeV scale) physics effects 
that we have discussed so far with the large distance physics (dark matter) effects.
We therefore analyze the dark matter constraints on CMSSM, whose LSP is a good 
dark matter candidate. To this purpose, one takes the  phase space points 
in Figure~\ref{dplot}, and evaluates for each of them the relic density, $\Omega h^2$, 
using micrOMEGAs \cite{Belanger:2006is} and  test if it is consistent with WMAP \cite{wmap}.
The result is presented in Figure~\ref{omcon}, where we
show the points consistent with the WMAP value as well as those 
that saturate it within  $3 \sigma$.  The plot is very similar for a 1$\sigma$ 
saturation  of WMAP value, with only minor differences.  
Requiring minimal $\Delta$ and consistency with WMAP
leads to the predictions:
\bea\label{mhfromrd}
m_h&=&114.7\pm 2 \,\,\,\mbox{GeV}, \,\,\,\,\,\Delta=15.0,\,\,\,\,\,
\textrm{(consistent with WMAP bound).}\nonumber\\
m_h&=& 116.0\pm 2 \,\,\,\mbox{GeV}, \,\,\,\,\,\Delta=19.1,\,\,\,\,\,
\textrm{(saturating the WMAP within 1$\sigma$).}\nonumber\\
m_h&=&115.9\pm 2 \,\,\,\mbox{GeV}, \,\,\,\,\,\Delta=17.8,\,\,\,\,\,
\textrm{(saturating the WMAP within 3$\sigma$).}
\eea

\smallskip
We checked  that  using a different definition for $\Delta$, such as 
$\Delta^\prime$ of (\ref{ss34}) does not change these results for 
$m_h$, since  $\min(\Delta^\prime)$ is found at similar values
for $m_h$ and its plot as a function of $m_h$ with its dark matter constraints
is indeed very similar (not shown here). 
The main difference is an overall shift of the fine-tuning plots of
Figure \ref{omcon} towards 
higher values of fine tuning, by a factor between 1.5 and 2. The minimum
of $\Delta^\prime$ indicates however the same value of $m_h$.

To conclude, minimizing the fine-tuning together with the constraints
from precision electroweak data, the bounds on Susy masses and  the requirement of the 
observed dark matter abundance lead to a prediction for $m_h$,
without imposing the LEP2 bound, that is marginally above this bound. 
This is an interesting result, and represents our prediction \cite{Cassel:2010px}
for the CMSSM lightest Higgs mass based on assuming $\Delta$ as a quantitative test of 
Susy as a solution to the hierarchy problem.

While minimal values of $\Delta$ are preferable, one can nevertheless ask what the
bounds on the parameter space are, for a  $\Delta$ beyond which supersymmetry 
is considered to fail to solve the hierarchy problem. It is in general agreed that
this happens  for  $\Delta\!\geq\!100$ or so. Therefore the bound $\Delta\!<\!100$ 
together with the dark matter consistency (3$\sigma$ upper limit), generates 
the following upper limits on the CMSSM parameters and $m_h$:
\medskip
\begin{eqnarray}\label{cmssmparameters}
m_{h}<~121~\mbox{GeV}, \qquad \mu<657.2~\mbox{GeV},\qquad m_0<3.2~\mbox{TeV} \qquad
\nonumber\\
127.6~\mbox{GeV}< m_{1/2}<~599~\mbox{GeV},\qquad
-1.76~\mbox{TeV}~< A_{0} <~2.26~\mbox{TeV}
\end{eqnarray}

\medskip\noindent
These values can be  re-evaluated for a different upper value  of $\Delta$.
Note the value of $m_h$ that corresponds to EW fine-tuning of 1 part in 100. 
Given the exponential increase of $\Delta$ with $m_h$, one sees that at $m_h=126$ GeV
one already  has a very large, unacceptable fine-tuning $\Delta=1000$.
The mass limits in Table~\ref{partlimit} scale approximately as 
$\sqrt{\Delta_{\mbox{\tiny min}}^{}}$, 
so they may be adapted depending on how much tuning one is willing to accept. 

\bigskip

\begin{table}[tbh]
\begin{center}
\begin{tabular}{|c||c|c|c|c||c|c||c|c||c|c|}
\hline
$\tilde{g}$ & $\chi_{1}^{0}$ & $\chi_{2}^{0}$ & $\chi_{3}^{0}$ & $
\chi_{4}^{0}$ & $\chi_{1}^{\pm}$ & $\chi_{2}^{\pm}$ & $\tilde{t}_{1}^{}$ & $
\tilde{t}_{2}^{}$ & $\tilde{b}_{1}^{}$ & $\tilde{b}_{2}^{}$ \\ \hline\hline
1720 & 305 & 550 & 660 & 665 & 550 & 670 & 2080 & 2660 & 2660 & 3140 \\ 
\hline
\end{tabular}
\end{center}
\vspace{-0.2cm}
\caption{\small 
Upper mass limits on superpartners (GeV) for which $\Delta<100$ 
(no dark matter constraint). 
If any of these states have masses larger than those shown, 
will require a fine-tuning worse than 1\%.}
\label{partlimit}
\end{table}
\medskip

\begin{figure}[th!]
\begin{center}
\begin{tabular}{cc|cr|} 
\parbox{6.9cm}{\psfig{figure=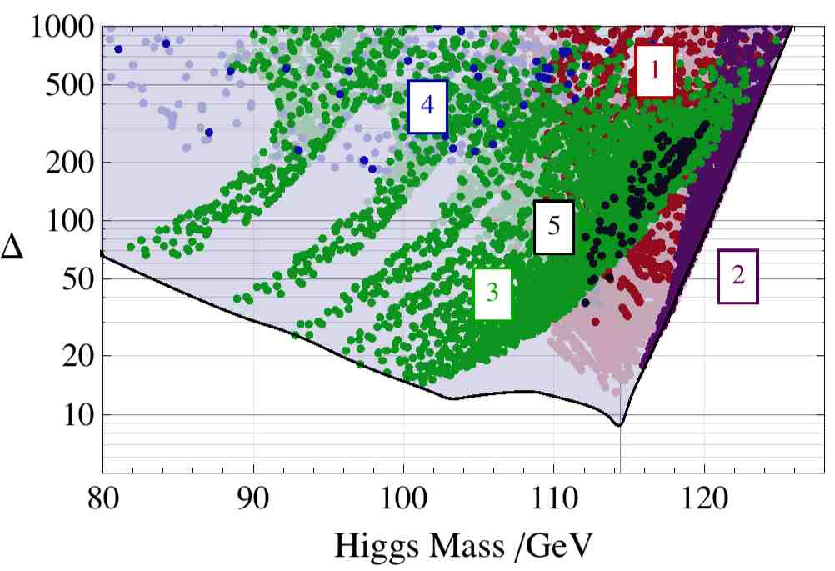,height=5.5cm,width=6.9cm}} 
\hspace{0.4cm}
\parbox{6.9cm}{\psfig{figure=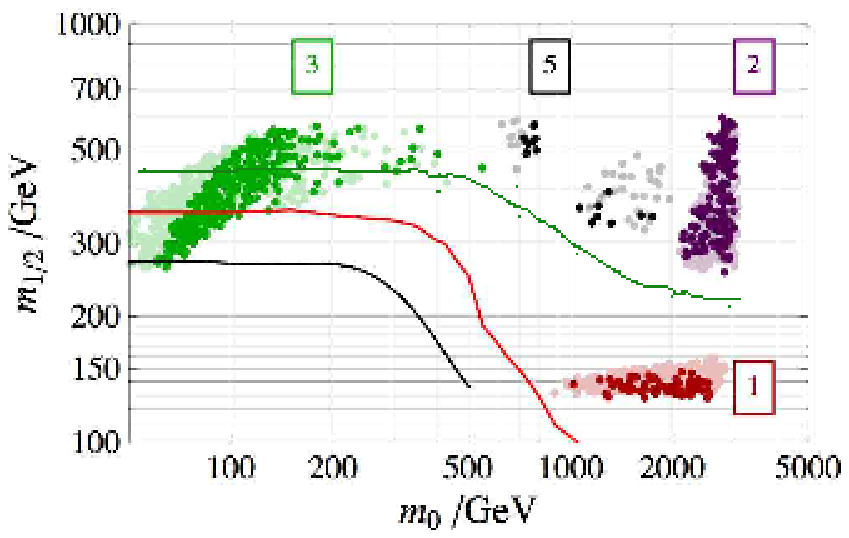,
height=5.8cm,width=6.9cm}} 
\end{tabular}
\end{center}\renewcommand{\baselinestretch}{1.}
\caption{\protect\small 
Left: Two-loop fine-tuning versus Higgs mass 
for the scan over CMSSM parameters with no constraint on the 
Higgs mass. This is the same plot as in figure~\ref{dplot} but with different
colour encoding.
The dark green, purple, red and black coloured regions have a 
dark matter density within 
$\Omega h^{2} = 0.1099 \pm 3 \times 0.0062$ 
while the lighter coloured versions of these regions lie below this 
bound $\Omega h^2<0.1285$ ($3\sigma$). 
The colours and associated numbers refer to different LSP 
structures: Regions 1, 3, 4 and 5 have 
an LSP which is mostly bino-like. In region~2, the LSP has a 
significant higgsino component, about $10\%$.
Right: Regions of fine tuning $\Delta<100$, summed over $\tan\beta$ and $A_0$ with
$m_h>111$ GeV (to account for the 2-3 GeV theoretical error). Same colour
encoding as in the left plot.  The CDMS-II bound \cite{Ahmed:2009zw}
is applied and reduces the area of  purple points.
The black, red and green curves correspond to exclusion curves from 
CMS, Atlas 2010 curve \cite{daCosta:2011qk} and Atlas 2011 curve
\cite{arXiv:1109.6572} respectively, see figure~\ref{omcon2atlas} for details.}
\label{oq2}
\end{figure}

It is interesting to compare the results in eq.(\ref{cmssmparameters}) for the parameter 
space of CMSSM with $\Delta\!<\!100$ and  dark matter consistency,
with the recent exclusion limits of the  Atlas/CMS experiments  
\cite{daCosta:2011qk, arXiv:1109.6572}, shown in 
Figure~\ref{omcon2atlas}. As it can be seen from 
Figures~\ref{omcon2}, \ref{oplots}, \ref{oq2},
LHC experiments already probe points in the  parameter 
space that have  fine-tuning $\Delta<100$. 
  Note however that the experimental 
exclusion curves  are for very specific values of the parameters 
($\tan\beta, A_0,....$) which are not summed over, and
therefore points below these curves that have different values for these 
parameters are not tested or ruled out.

The situation changes for different values of $\Delta$  that one is prepared to accept, 
see the plots in Figures~\ref{oplots} together with the Atlas/CMS exclusion limits.  
In particular the case with $\Delta<20$ selects
points with the smallest fine tuning. It should be stressed that the presence
of fewer points (smaller, shrinking area) in the ``moduli'' space ($m_0,m_{1/2}$) 
does not imply  anything special for the overall fine-tuning amount, like
a larger $\Delta$ (disproved by Figures~\ref{oplots}). It can for example be related to the 
quality of the scan of the parameter space. Apart from this, it  
indicates the most likely values of the parameter (moduli) space that 
are preferred by the low energy physics, that a fundamental theory (like a string model)
should explain or fix dynamically. For 
related interpretations of the recent LHC 
results see  \cite{Allanach:2011wi,Buchmueller:2011aa}.

It is important to note that most of the region with $\Delta<100$  of 
the parameter space is being tested by the combined LHC (at 7 TeV) 
and CDMS-II and Xenon experiments \cite{Ahmed:2009zw},  as it was discussed in 
 \cite{Cassel:2011tg} together with the possible 
detection signals. Here we only make some comments on the impact of 
LHC and dark matter experiments on the various regions of the fine-tuning plots
shown so far. In figure~\ref{oq2}
colour encoding corresponds to the structure of the LSP and shows where 
points of different $(\Delta, m_h)$ are located in the plane $(m_0,m_{1/2})$.
Red points (1) have a low gluino mass, the LSP is mostly bino with a cross section 
off nuclei too 
small to be probed by the next generation of direct dark matter searches. Green points (3)
have lighter squarks, mostly bino LSP 
and together with the red points are reachable by
 LHC run I (right plot).  Black points (5) have gluino and squarks near the TeV scale
and an almost pure bino LSP.  Purple points (2)  
have a significant higgsino component ($10\%$),  a heavy gluino, ($\sim 900$  GeV or larger) 
and TeV-scale squarks.  Although  they are not within the reach of LHC run I, they
are sensitive to   direct dark matter searches; the  CDMS II bound
is applied in their case while Xenon100 (not applied)
can reduce their area further \cite{Ahmed:2009zw}.
Also, as seen from the left plot the purple region includes points of
very low fine-tuning. In conclusion
dark matter and LHC run I searches are rather complementary in covering the entire
plane $(m_0,m_{1/2})$ in the TeV region. To cover all the parameter space, beyond $\Delta<100$
will require running at the full 14 TeV CM energy.

\section{The impact of ``new physics'' beyond MSSM on $\Delta$ and  $m_h$.}

Following this analysis, some natural questions emerge.
How do the above predictions for $m_h$ and $\Delta$ change
under the presence of 
``new physics'' that  may exist at some high scale (few TeV or so), 
beyond the MSSM Higgs sector? 
Recall that the MSSM Higgs potential is the most minimal construction allowed
by supersymmetry, but new physics in this sector can exist: for example 
extra Higgs doublets,  gauge singlets, additional massive U(1)$^\prime$ 
bosons, which can all affect the MSSM Higgs sector, its
quartic effective coupling and its predictions.
Another question is the following: assume for a moment that a 
Higgs particle is not found at the minimal $\Delta$ as discussed above, 
is it then possible to have a larger $m_h$, of say 121 GeV but with 
$\Delta\sim \cO(10)$ instead of the corresponding value found above 
of $\Delta\!=\!100$?

One answer is  that the model considered is  too constrained, and this may
lead to a large $\Delta$. Indeed, it is known that 
gaugino universality condition considered here, if 
relaxed, reduces the value of $\Delta$ \cite{DG}; 
another possibility is to relax the universal Higgs mass
which can also reduce $\Delta$. Finally, another possibility is 
that  ``new physics'' missed by the CMSSM Higgs sector, 
can increase  the effective quartic Higgs coupling which 
leads to a reduction of the fine-tuning amount.
The ways to achieve this are  numerous but also model dependent. 

A simple possibility is to consider the case of MSSM with a 
low supersymmetry breaking scale ($\sqrt f\sim$ few TeV ) in the hidden sector
\cite{Navarro,arXiv:1006.1662}.
In this case, when integrating out the auxiliary field of the 
goldstino superfield that is coupled to the MSSM,  one generates
apart from  the usual soft terms, additional quartic terms,  without
introducing new parameters in the visible sector of the model. 
Indeed, the scalar potential contains a term \cite{arXiv:1006.1662}
\medskip
\bea
V \supset \frac{1}{f^2}\,\big\vert \,m_1^2\,\vert h_1\vert^2+m_2^2\,\vert h_2\vert^2
+B_0\,h_1.h_2\, \big\vert^2
\eea

\medskip\noindent
in addition to the MSSM potential, in the standard notation. This term 
can be significant for low $f$ while in the more familiar cases of 
high scale Susy breaking is strongly suppressed. Such term
can help to increase the SM-like Higgs mass at the tree level and reduce
$\Delta$ since 
\medskip
\bea
\Delta_p\propto \frac{p}{2\,v^2 \,m_2^4/f^2+(1+\delta)\,m_Z^2}+\cO(1/\tan\beta)
\eea

\medskip\noindent
can be reduced compared to its MSSM counterpart in eq.(\ref{fggf}), due
to the extra term in the denominator. For more
details of this class of models see \cite{arXiv:1006.1662}  and references therein.

Other possibilities to increase $m_h$ and reduce $\Delta$ can exist, due to ``new physics''.
To perform a model independent analysis, below 
we consider an effective theory approach, that parametrizes 
in  a general way the ``new physics'' that may exist beyond the MSSM Higgs sector.
This is done by using a series of effective operators, whose consequences 
for $\Delta$ and $m_h$ 
are explored below. Let us mention  that introducing 
such operators comes at the cost of having new 
parameters in the theory, beyond those of the MSSM.

The power of such an effective  approach resides in its organizing principle 
that relies  on an expansion in inverse powers of the scale $M$ of the "new physics" 
that generated these operators. 
To this purpose we consider the  effective operators of dimensions d=5 and d=6 that can exist
in the Higgs sector.  The reason of considering both classes is twofold:
first,  the  latter can be present as a leading contribution even in the absence of the former.
For example integrating out a massive U(1)$^\prime$ generates  d=6 operators, with no
d=5 ones. The second reason is that, if generated by the same physics, at large $\tan\beta$ 
the d=5 operators receive additional suppression and then become of similar order of 
magnitude to the d=6 operators. In other words, the convergence of the expansion (in $1/M$) 
for large $\tan\beta$ requires one include both classes of operators, when originating 
from same massive state that was integrated out.

\subsection{The case of d=5 operators and their corrections to $m_h$.}

We start with the d=5 operators in the MSSM Higgs  sector. There are two of them:
\medskip  
\begin{eqnarray} 
\mathcal{L}_{1} &=&\!\!\frac{1}{M}\int d^{2}\theta \,
\,\lambda'_H(S)\,(H_{2}.H_{1})^{2}\!+\!h.c.,
\quad
\lambda'_H(S)/M
= \zeta_{10}+ \zeta_{11}\,S,\,\,
\quad (\zeta_{10},\, \zeta_{11}\sim 1/M),
\nonumber\\
\mathcal{L}_{2} &=&\!\!\frac{1}{M}\int d^{4}\theta \,\,\Big\{%
\,a(S,S^{\dagger })
D^{\alpha }\Big[b(S,S^{\dagger })\,H_{2}\,e^{-V_{1}}\Big]%
D_{\alpha }\Big[c(S,S^{\dagger })\,e^{V_{1}}\,H_{1}\Big] 
+h.c.\Big\} \label{dimensionfive}
\end{eqnarray}

\medskip\noindent
where $S=\theta\theta m_0$ is the spurion field and we assume $m_0\ll M$ 
so that the expansion is convergent. Here $a,b,c$ are all dimensionless
functions of the spurion
that have the form $a(S,S^\dagger)=a_0 +a_1 S+a_1^*\,S^\dagger+a_2 S\,S^\dagger$, and similar 
for $b$ and $c$.

$\cL_1$ can be generated by integrating out a
massive gauge singlet or $SU(2)$ triplet.
Let us consider the former case. 
Indeed, in the MSSM with a massive gauge singlet, with 
an F-term $M \Sigma^2+\Sigma\,H_1.H_2$,
when integrating out  $\Sigma$  via the eqs of motion, one
generates $\cL_1$. This  result is similar to considering a 
generalised NMSSM with a  supersymmetric mass term $M \Sigma^2$, in 
the decoupling limit of the singlet. Therefore our results
for fine tuning are relevant for  the generalised NMSSM model,
in this limit.
 
Regarding $\cL_2$, this can be generated in various ways (see Appendix A,
 B in \cite{Antoniadis:2008es}) but perhaps the
simplest way is via an additional pair of massive Higgs doublets of
mass of order $M$. However, $\cL_2$ is actually "redundant", since it can be removed
by general spurion-dependent field redefinitions, up to  soft terms renormalisation,
$\mu$ term redefinition and $\cO(1/M^2)$ corrections \cite{Antoniadis:2008es}.
Therefore $\cL_2$ plays no role in the following. Note that $\cL_2$ also includes
a particular $d=5$ operator of the form $\int d^2\theta H_1\Box H_2$ which can be
re-written as a Kahler term (for details
see  \cite{Antoniadis:2007xc} and the Appendix  in the first 
work in \cite{Antoniadis:2009rn}).

\begin{figure}[t] 
\begin{center}
\begin{tabular}{cc|cr|} 
$\Delta$\parbox{6.5cm}{ 
\psfig{figure=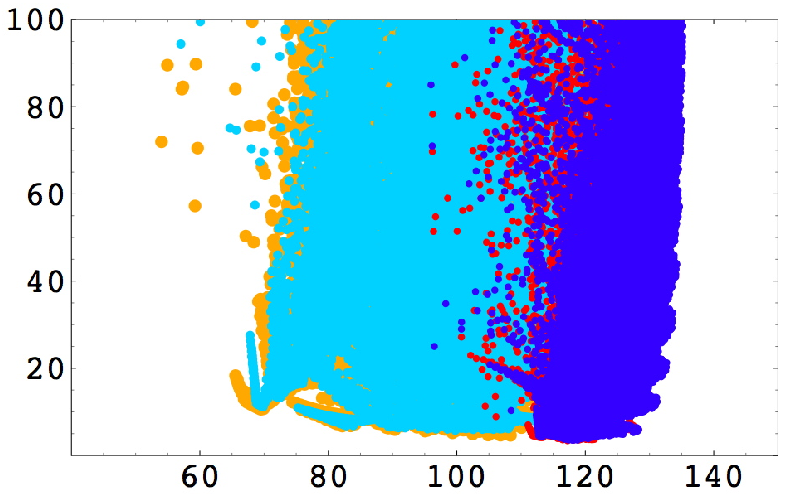,  
height=4.5cm,width=6.1cm}
} 
\hspace{0.4cm}  
$\Delta$\parbox{6.5cm}{ 
\psfig{figure=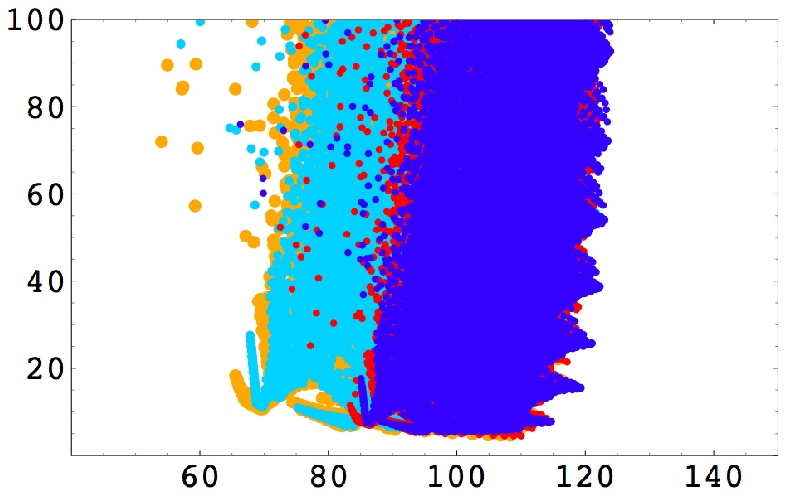, 
height=4.5cm,width=6.1cm}
} 
\end{tabular}
\end{center}
\vspace{-0.1cm}
\hspace{4cm}$m_h$\hspace{7cm}$m_h$
\vspace{0.25cm}
\def\baselinestretch{1.07}
\caption{\protect\small  
Left figure (a): 
the fine tuning $\Delta $ as a function of $m_{h}$ (GeV) at one-loop;
$\Delta $ of MSSM is plotted in light blue ($m_{t}=174.4$ GeV) with an orange 
edge (shift induced for $m_{t}=171.8$ GeV) and extends up to $m_h\approx 114$ GeV 
from where it grows exponentially; 
$\Delta $ of MSSM with d=5 operators with
$(2\,\mu_0\,\zeta_{10})=0.07$, $(2 \,m_0\,\zeta_{11})=0$ 
is plotted in dark blue ($m_{t}=174.4$ 
GeV) and with a red edge (if $m_{t}=171.8$). 
Right figure (b): 
similar to figure (a) but with $\zeta_{10}=0$, $(2\,m_0\,\zeta_{11})=-0.1$. Non-zero or 
larger $\protect\zeta _{10,11}$ (dark blue and red areas) shift the plots to 
higher $m_{h}$, for fixed $\Delta$.
Similar behaviour is present for simultaneous non-zero value for both
coefficients $\zeta_i$.} 
\label{fig4ab} 
\end{figure} 

Due to $\cL_1$ 
the scalar potential acquires additional terms which 
bring a  correction to the effective quartic coupling:
$\lambda\rightarrow \lambda+ (2\zeta_{10}\mu_0)\,\sin 2\beta
+ (-1/2) m_0\zeta_{11}\,\sin^2 2\beta$. 
So $\lambda$ may increase and as  a result we 
expect that fine-tuning can be reduced. The correction to the MSSM 
lightest Higgs mass is then
\medskip
\bea\label{mhold}
m_{h,H}^2&=&(m_{h,H}^2)_{\textrm{MSSM}}
\nonumber\\
&+&
{(2\,\zeta_{10}\,\mu_0)\,
{  v}^2\,\sin 2\beta}
\,\Big[1\pm\frac{m_A^2+m_Z^2}{\sqrt{\tilde w}}\Big]+
\frac{(-2\,\zeta_{11}\,m_0)\,{  v}^2}{2}\,\Big[1\mp
\frac{(m_A^2-m_Z^2)\,\cos^2 2\beta}{\sqrt{\tilde w}}\Big]
\nonumber\\
&+&
\delta m_{h,H}^2, \qquad \textrm{with}\qquad \delta m_{h,H}^2
= \cO(1/M^2)
\eea
\medskip
and
\medskip
\bea\label{ma}
m_A^2=
(m_A^2)_{\textrm{MSSM}}-  
\frac{2\,\zeta_{10}\,\mu_0\,v^2}{\sin2\beta}
+2\,m_0\,\zeta_{11}\,v^2+\delta m_A^2,\quad \delta m_A^2=\cO(1/M^2)
\eea

\bigskip\noindent
for the pseudoscalar Higgs.  $\delta m_{h,H}^2$ and  $\delta m_A^2$ are  
$\cO(1/M^2)$ due to  $d=5$ and, if present, also  $d=6$ operators.
The upper (lower) signs correspond to $h$ ($H$),
and $\tilde w$ is given by
$\tilde w\equiv (m_A^2+m_Z^2)^2-4\,m_A^2\,m_Z^2\,\cos^2 2\beta$.
With this result one can show that the mass $m_h$ can be increased
above the LEP2 bound, for low $\tan\beta$, 
 also with the help of quantum corrections.

Let us now see the amount of fine-tuning as a function of the Higgs mass, in
the presence of d=5 operators\footnote{There is a subtle point: notice
that the RG eqs are not affected at one loop
by the presence of a massive gauge singlet that was assumed to generate 
$\cL_1$ in the first instance. Therefore this analysis is consistent.
The situation is more complicated in the presence of SU(2) triplet; however, 
the Higgs mass increase and the reduction of  fine-tuning due to it remain true, 
even though the exact value of $\Delta$ may be different.}.
Unlike the MSSM case the analysis of $\Delta$ is done 
at one-loop only, for a sample of points in parameter space with:
$1.5\leq \tan \beta 
\leq 10$, $50\,\mathrm{GeV}\leq m_{0},m_{12}\leq 1$ TeV,
 $130$~GeV~$\leq \mu_0\leq 1$ TeV,
$-10\leq A_{t}\leq 10$  and $171.8\leq m_{t}\leq 174.4$ GeV, consistent  
 with the signs for $\zeta_{10}, \zeta_{11}$ chosen so as to reduce the fine tuning. 
One-loop analytical results for $\Delta$ were obtained (for details see
\cite{Cassel:2009ps}), 
and the corresponding numerical results are shown in Figure~\ref{fig4ab}. 
Note that in these figures the structure apparent at small $\Delta$ and large $m_h$  is
a scanning artifact, with the under-dense wedge shaped
 regions to be filled in with a more dense parameter sample.
For a fixed value of $\Delta$ and relative to the CMSSM case, 
one may see a systematic shift towards higher $m_{h}$.
This is  due to an increase of the effective quartic Higgs coupling (which 
also increases $m_h$ relative to its MSSM value) and, as a result, also 
decreases $\Delta$.  
The overall result is that the  minimum amount of fine-tuning $\Delta $ in the 
presence of $d=5$ effective operators  is reduced relative to the MSSM case,
so values of $m_{h}$ as large as $130$  GeV can still have a low $\Delta <10$. 
The reduction in the fine tuning at low $\tan\beta$ ($< 10$) relative 
to the MSSM case is actually much more  marked than that shown, given that 
in the MSSM, in this limit  and $m_h$ above the LEP2 bound, $\Delta$ actually 
increases.

Let us examine the scale of new physics needed for this reduction in fine 
tuning. One has
\medskip
\begin{equation}
M\approx \frac{1}{\zeta_{10}}=\frac{2\mu_0}{0.07}
\approx 30\, \mu_0,
\end{equation}

\bigskip\noindent
With $\mu_0$  between the EW scale and 1 TeV, 
this shows that large values of $M$ are allowed: $M\approx 6\, (9)$ TeV 
for $\mu_0=200 (300)$ GeV, respectively. Also, in this case
$\Delta<10$ for  $114\leq m_h\leq 130$ GeV.
To relax these values one can use that an increase (decrease) 
of $(2\,\mu_0\,\zeta_{10})$ by $0.01$ increases (decreases)
$m_h$ by $2$ to $4$ GeV for the same $\Delta$.

Let us conclude with a remark on fine-tuning in the presence of a massive
gauge singlet that was assumed to generate the above $d=5$ operator,  and a
 comparison to the NMSSM case. The above case is less fine tuned than the NMSSM
because of the nature of the 
{\it supersymmetric} contribution of the gauge singlet to effective 
quartic higgs coupling. In our case  $\lambda$ receives a contribution
$\lambda\sim \sin 2\beta$ (see eqs.(\ref{mhold})), while in the NMSSM the corresponding
contribution of the singlet 
is further  suppressed, being proportional 
to $\lambda\sim \sin^2 2\beta$.
As a result the NMSSM  is more fine-tuned (for a similar $m_h\propto 2 \lambda v^2$).
This difference can be further traced back to the absence in NMSSM of 
a bilinear F-term $M \Sigma^2$, that could increase the mass of the singlet significantly
(and generated the d=5 operator). Correspondingly, the increase of  
$\lambda$ and thus the reduction of the  fine-tuning in NMSSM  is 
not as significant  as in the case discussed above (for a fixed  $m_h$).

\subsection{The case of d=6 operators and their correction to $m_h$.}

We can extend the previous discussion to also include the corrections to $m_h$ from
 effective operators of dimension $d=6$  that can exist beyond  the MSSM Higgs sector.
Computing the analytical corrections to $m_h$ from individual effective operators
is indeed possible. It is however
difficult to evaluate the fine tuning $\Delta$ in this case, since the operators can
be generated in various ways by (unknown) states charged under the SM group, 
that affect the RG flow of couplings in the model 
and thus the fine tuning. This is unlike the case of $\cL_1$ in the $d=5$ case which
was assumed to be generated by a massive gauge singlet. 
Nevertheless, it is obvious that an increase
of $m_h$ by the effective operators reduces the fine tuning $\Delta$.
For these reasons, in the following we restrict ourselves to computing 
the corrections to $m_h$.

The list of $d=6$ effective operators which are polynomial in fields is  
\cite{Cassel:2009ps,Piriz:1997id}
\medskip
\bea
&&\!\!\!\!\!\!
\mathcal{O}_{1} =
\frac{1}{M^{2}}\int d^{4}\theta\,\mathcal{Z}_{1}\,(H_{1}^{\dagger }\,e^{V_{1}}\,H_{1})^{2}, 
\qquad\qquad\quad\,\,
\mathcal{O}_{5}=\frac{1}{M^{2}}\int\! d^{4}\theta 
\mathcal{Z}_{5} (H_{1}^{\dagger}\,e^{V_{1}}\,H_{1})\,\,H_{2}.\,H_{1}+h.c.
\nonumber\\[-2pt]
&&\!\!\!\!\!\!\mathcal{O}_{2} =
\frac{1}{M^{2}}\int\! d^{4}\theta \,
\mathcal{Z}_{2}\,
(H_{2}^{\dagger }\,e^{V_{2}}\,H_{2})^{2}, 
\qquad\qquad\quad\,\,
\mathcal{O}_{6}=
\frac{1}{M^{2}}\int\! d^{4}\theta \,\mathcal{Z}_{6}\,
(H_{2}^{\dagger }\,e^{V_{2}}\,H_{2})\,\,H_{2}.\,H_{1}+h.c. 
\nonumber\\
&&\!\!\!\!\!\!\!
\mathcal{O}_{3} =
\frac{1}{M^{2}}\int\! d^{4}\theta \,\mathcal{Z} _{3}
(H_{1}^{\dagger }\,e^{V_{1}}\,H_{1})\,(H_{2}^{\dagger }
\,e^{V_{2}}\,H_{2}),
\quad
\mathcal{O}_{7}=\frac{1}{M^{2}}\int\! d^{2}\theta \,
{\mathcal{Z}_{7}}\,{\rm Tr}\,
W^{\alpha }\,W_{\alpha }\,(H_{2}\,H_{1})+h.c.
\nonumber\\
&&\!\!\!\!\!\!
\mathcal{O}_{4}=\frac{1}{M^{2}}\int\! d^{4}\theta \,
\mathcal{Z}_{4}\,(H_{2}.\,H_{1})\,(H_{2}.\,H_{1})^{\dagger },
\qquad\,\,\,\,
\mathcal{O}_{8}=\frac{1}{M^{2}}\int\! d^{4}\theta
\mathcal{Z}_{8}\,\,(H_{2}\,H_{1})^{2}+h.c.\qquad
\label{operators18}
\eea

\medskip\noindent
where $W^\alpha=(-1/4)\,\overline D^2 e^{-V} D^\alpha\, e^V$
is the chiral field strength
of $SU(2)_L$ or $U(1)_Y$ vector superfields $V_w$ and $V_Y$ respectively.
 Also  $V_{1,2}=V_w^a
(\sigma^a/2)+(\mp 1/2)\,V_Y$ with the upper sign for $V_1$.
Finally, the wavefunction coefficients are spurion dependent and
have the structure
\medskip
\bea\label{z1}
({1}/{M^2})\,\cZ_i(S,S^\dagger)=\alpha_{i0}
+\alpha_{i1}\,S
+\alpha_{i1}^*\,m_0\,S\,S^\dagger
+\alpha_{i2}\,m_0^2\,S\,S^\dagger,\qquad
\alpha_{ij}\sim 1/M^2.
\eea

\medskip\noindent
where $S=m_0\theta\theta$.  $\cO_{1,2,3}$ can be
generated in the MSSM with an additional,
massive $U(1)'$ gauge boson or $SU(2)$ triplets, when
these are integrated out \cite{Dine}.
$\cO_4$ can be generated by a massive gauge singlet or $SU(2)$
 triplet,  while $\cO_{5,6}$
can be generated by a combination of $SU(2)$ doublets
and massive gauge singlet. $\cO_7$ is essentially a 
threshold correction to the gauge coupling, with a 
moduli field replaced by the Higgs,  difficult to generate in a 
renormalisable theory with additional massive states.  Finally, $\cO_8$ 
exists only in the non-Susy case, but is generated when removing
the $d=5$ derivative operator $\cL_2$ by field redefinitions \cite{Antoniadis:2008es}, 
therefore we keep it. There are also operators which involve derivatives (see later).

\begin{figure}[t!] 
\begin{center}
\begin{tabular}{cc|cr|} 
\parbox{6.5cm}{\psfig{figure=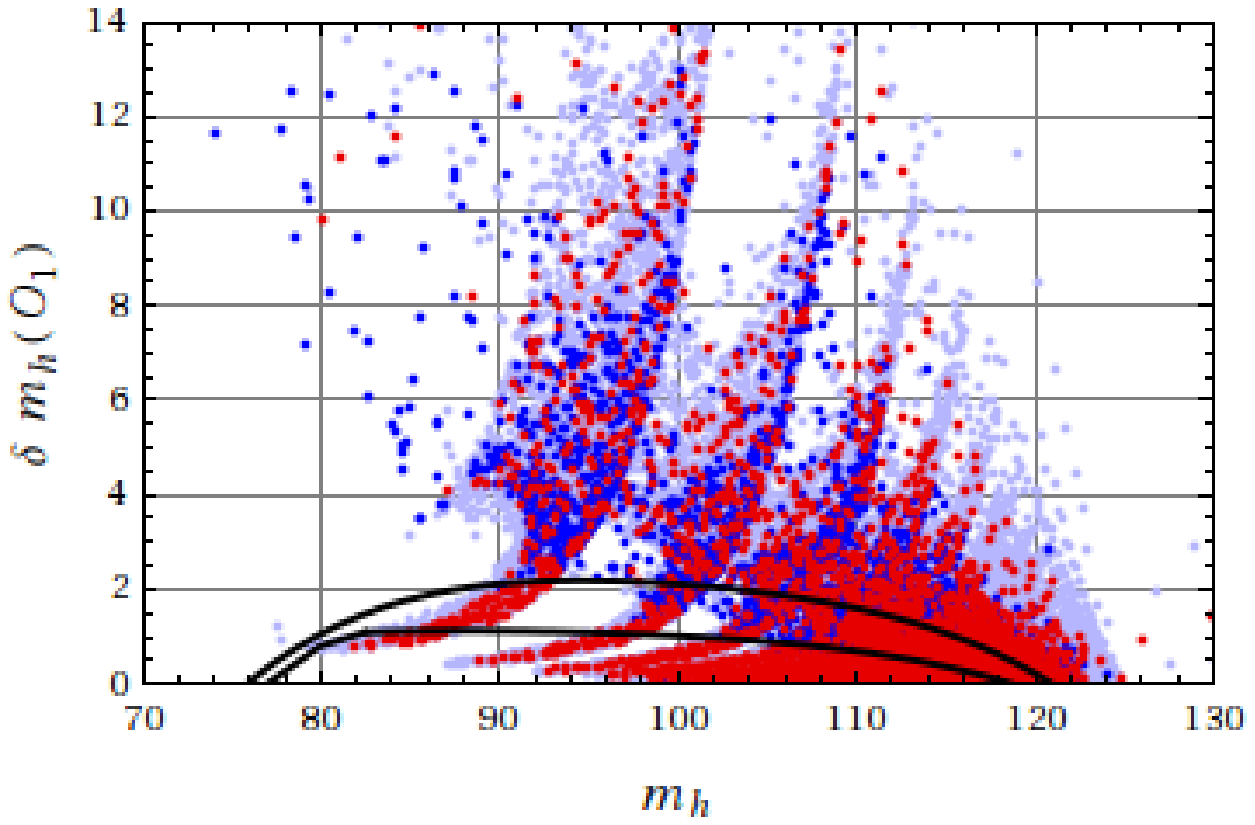, height=5.cm,width=6.4cm}} 
\hspace{0.4cm}  
\parbox{6.5cm}{\psfig{figure=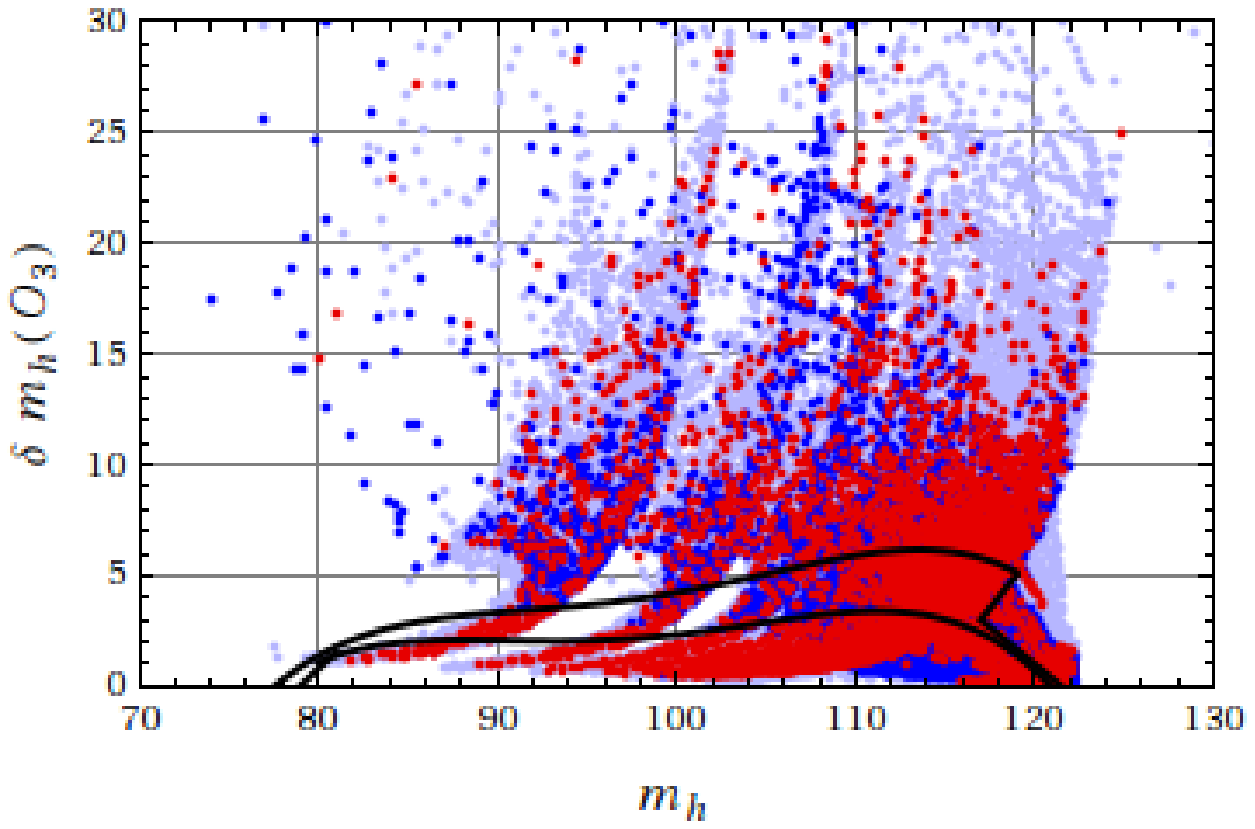, height=5.cm,width=6.4cm}}
\end{tabular}%
\end{center}
\def\baselinestretch{1.}
\caption{\small
The corrections $\delta m_h$ due to  $\cO_1$ (left plot) and $\cO_3$ (right plot) operators, 
in  function of the 2-loop (leading-log) $m_h$ of the CMSSM, with $M=8$ TeV. 
Only the Susy part of these operators is considered, labelled by 
$\alpha_{10}$ ($\alpha_{30}$), respectively. Light blue: CMSSM 
points with relic density  $\Omega h^2\geq 0.1285$; on top, in dark blue are
CMSSM points with $\Omega h^2\leq 0.0913$ (3$\sigma$ deviation); on top,
in red, are CMSSM points that saturate WMAP bound within 3$\sigma$ (WMAP value
$\Omega h^2=0.1099\pm 0.0062$). The total, corrected value of
 Higgs mass is then $m_h+\delta m_h$. 
The CMSSM points below the lower continuous line have $\Delta\!<\!100$, those between 
the two continuous lines have ($100\!\leq\! \Delta\!\leq\!200$)
while those above  the upper continuous line have $\Delta\!>\!200$.
Therefore $\delta m_h$ for $\Delta<100$ is rather small, of few GeV. 
$\cO_2$ ($\cO_4$) gives results remarkably close to 
those of $\cO_1$ ($\cO_3$), respectively, and are not shown.} 
\label{f1} 
\bigskip\medskip
\begin{center}
\begin{tabular}{cc|cr|} 
\parbox{6.5cm}{\psfig{figure=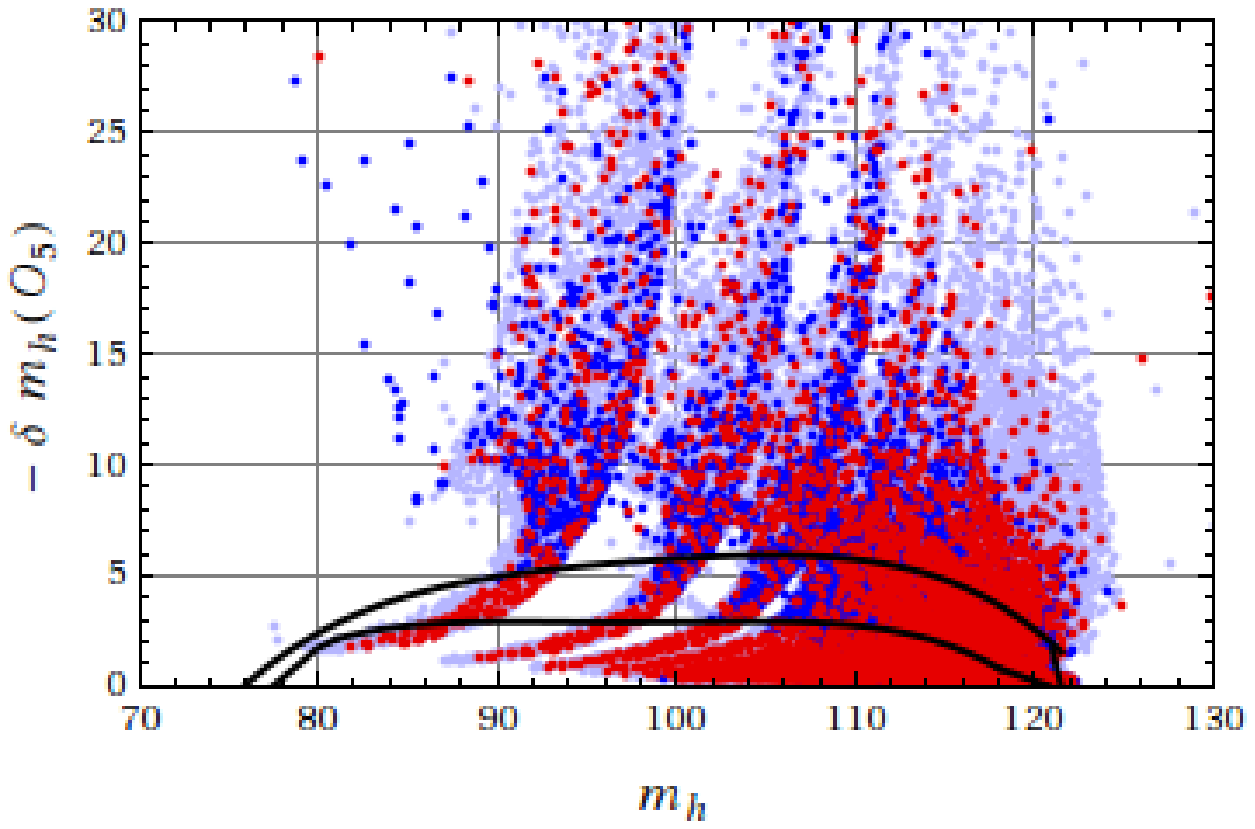,  
height=5.cm,width=6.4cm}} \hspace{0.4cm}  
\parbox{6.5cm}{\psfig{figure=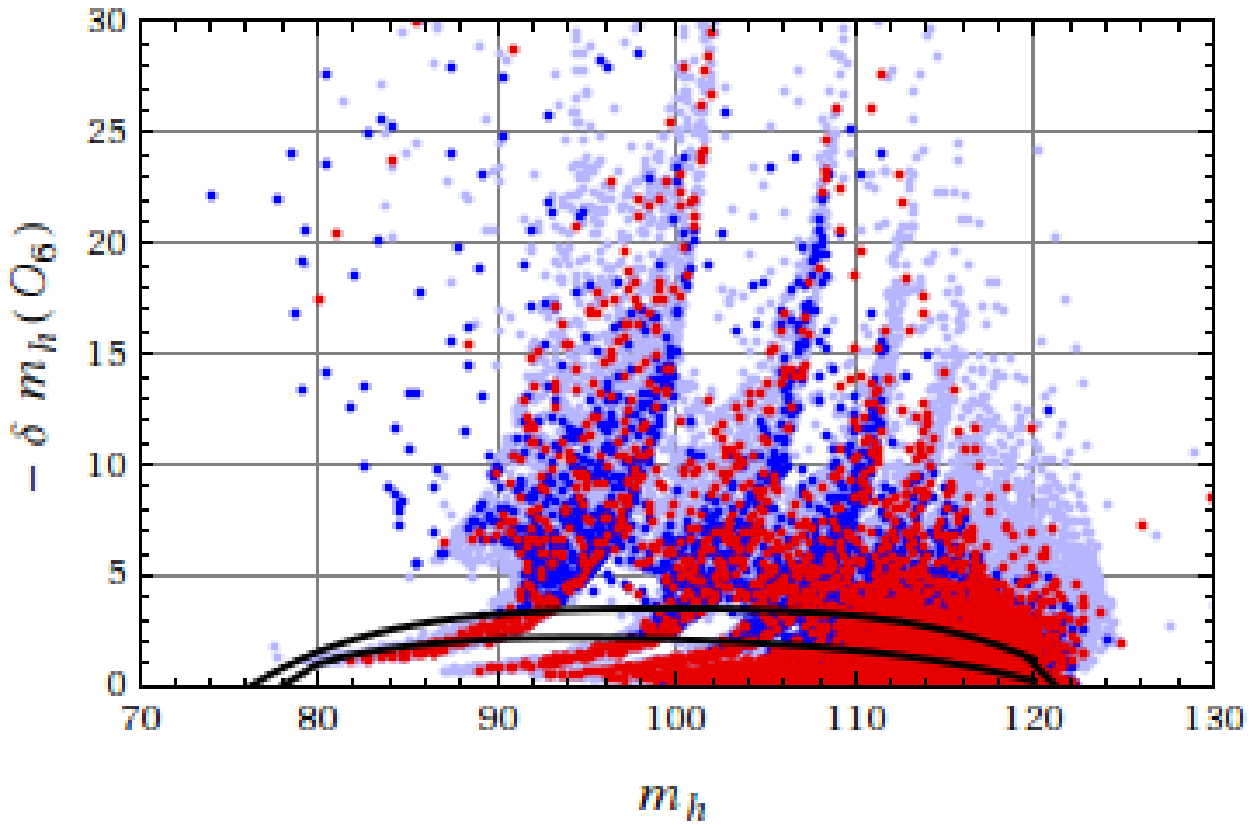, 
height=5.cm,width=6.4cm}}
\end{tabular}%
\end{center} 
\def\baselinestretch{1.}
\caption{\small
As for Figure~\ref{f1}, but for the supersymmetric part of
 $\cO_{5}$ (left) and $\cO_6$ (right) operators. Note the minus sign in
front of $\delta m_h$ showing that for the chosen (positive) sign of parameters
($\alpha_{j0}$) a decrease of $m_h$ is actually obtained. } 
\label{f2} 
\end{figure} 

Let us see the implications of these operators for the corrections to
the Higgs spectrum. They bring  $\cO(1/M^2)$ corrections denoted $\delta m_{h,H}^2, \delta
m_A^2$ in eq.(\ref{mhold}), (\ref{ma}) and their 
exact expressions can be found in \cite{Antoniadis:2009rn,Carena:2009gx}.
However, for most purposes, an expansion of these in $1/\tan\beta$
is accurate enough. At large $\tan\beta$,  
$d=6$ operators  bring corrections comparable to those of
 $d=5$ operators.  The relative $\tan\beta$  enhancement of $\cO(1/M^2)$ 
corrections compensates for their extra  suppression  relative 
to $\cO(1/M)$ operators. However, in some models
 only $d=6$ operators may be present, depending on the details of
the ``new physics'' generating the effective operators. 
If $m_A$ is kept fixed, one finds in $\cO(1/M^2)$ order
\cite{Antoniadis:2009rn} (see also \cite{Carena:2009gx}):
\medskip
\bea\label{dd1}
\delta m_{h}^2
\!\!\!&=&\!\!
-2\, v^2\,\Big[ \alpha_{22} \,m_0^2+(\alpha_{30}+\alpha_{40})\mu_0^2
+2 \alpha_{61} \,m_0\,\mu_0 
- \!\alpha_{20}\,m_Z^2\Big]
-\!
(2\,\zeta_{10}\,\mu_0)^2\,v^4\,(m_A^2-m_Z^2)^{-1}
\nonumber\\
&+&\!\!\!\! \!{v^2}{\cot\beta}
\Big[(m_A^2\!-\! m_Z^2)^{-1}
\!\Big( 4 \,m_A^2\,\big(\,
(2 \alpha_{21}\!+\!\alpha_{31}\!+\!\alpha_{41}\! +\!2 \alpha_{81})
m_0\,\mu_0\! +\! (2\alpha_{50}\! +\!\alpha_{60})\,\mu_0^2
+\alpha_{62} m_0^2\big)
\nonumber\\
&-&\! (2 \alpha_{60}\!-\!3\alpha_{70})\,m_A^2\,m_Z^2
\!-\!(2\alpha_{60}\!+\!\alpha_{70})\,m_Z^4\Big)
+{8 (m_A^2\!+\!m_Z^2)(\mu_0 m_0 \zeta_{10}\zeta_{11})
\,v^2}/{(m_A^2\!-\!m_Z^2)^2}
\Big]\nonumber\\[2pt]
&+&\cO(1/\tan^2\beta)
\eea

\medskip\noindent
The full value  of $m_{h}^2$ is that obtained by using eq.(\ref{dd1}) in
(\ref{mhold}) where $(m_h^2)_{MSSM}$ is replaced by the 2-loop 
leading log value\footnote{In the numerical evaluations and figures 
below we used the {\it exact} expression of $\delta m_h^2$
\cite{Antoniadis:2009rn}, see also \cite{Carena:2009gx}.}.
So the effective operators correction is regarded as a classical correction
 (``perturbation'') added to the 2-loop leading-log CMSSM value, and crossed-terms that involve
products of loop-corrections and effective operators coefficients are 
neglected, being of higher order. For an easier reading of the effects 
of the operators, one can introduce the correction $\delta m_h$
\medskip
\bea
\delta m_h=\Big[
m_h^2\big\vert_{\rm 2-loop,{\rm CMSSM}}+\delta m_h^2\Big]^{1/2}\!\!\!\!
-m_h\big\vert_{\rm 2-loop,{\rm CMSSM}}
=\frac{1}{2} \frac{\delta m_h^2}{m_h\big\vert_{\rm 2-loop,{\rm CMSSM}}}+\cO(1/M^4)
\eea

\medskip\noindent
with $\delta m_h^2$ as in (\ref{dd1}).
The total Higgs mass  value is then $\delta m_h+m_h\vert_{\rm 2-loop,CMSSM}$.
Further, it is preferable to search for possible increases of 
$m_h^2$ by supersymmetric  rather than supersymmetry-breaking effects
of the effective operators,  because the latter are less under control
in the effective approach\footnote{Also, one would prefer a {\it supersymmetric}
solution to the  fine-tuning problem associated with increasing
the MSSM Higgs mass well above the LEP2 bound.}. 
In any case, the non-Susy part of the effective operators has an 
impact on $\delta m_h$ that is  comparable to that of the supersymmetric 
part considered here. So in the following we concentrate  only on
Susy corrections,  induced by the coefficients $\alpha_{j0}$
with $j$ to label the corresponding operator $\cO_j$.

The results obtained are illustrated in Figures~\ref{f1}, \ref{f2}
where the correction $\delta m_h$ is shown as a 
function of the CMSSM value of $m_h$ (2-loop, leading-log). The results are obtained in the
the following way. We consider all the phase space points displayed 
in Figure~\ref{omcon} 
that respect all experimental and dark matter constraints (except the LEP2
 bound on $m_h$ that is not imposed). 
On this ``background'' we consider the {\it perturbation}
 due to the effective operators of dimension 
d=6  and evaluate $\delta m_h$ as outlined above and shown in these figures, 
as a function of the  2-loop leading-log value of $m_h$ in  CMSSM.

From these figures one can conclude that the CMSSM points with lowest fine-tuning
$\Delta\!<\!100$ corresponding to  $m_h\!<\!121$ GeV, have a rather 
small $\delta m_h$, of few GeV. For $M=8$ TeV $\delta m_h$ is up to 4 GeV  
($6$ GeV for $\Delta<200$), and this decreases  (increases) by $\approx 1$ 
GeV for a  1 TeV increase (decrease) of $M$. However, the value of $M$ is usually 
restricted to be in the region of $8$ TeV or higher, from $\rho$ parameter 
constraints\footnote{This  bound applies to a combination 
of operators and can in principle be reduced for individual operators.} \cite{blum} 
and our expansion parameter $\tilde m^2 \alpha_{ij}$ was always taken $<1/4$, where 
$\tilde m$ is any scale of the model ($m_0$, $m_{1/2}$, $\mu$). This is considered  
conservative enough for a convergent perturbative expansion. Note that points which 
were below the LEP2 bound by the  correction $\delta m_h$ mentioned, 
become now phenomenologically viable. The special point of CMSSM of minimal
$\Delta = 17.8$ that saturates the relic density within 
3$\sigma$  and with $m_h = 115.9\pm 2$ GeV, could receive a correction 
$\delta m_h \sim 4$  GeV, so that $m_h$  increases to $m_h+\delta m_h = 119.9\pm 2$ GeV. 
Correspondingly, its earlier associated  $\Delta$ is likely to decrease
by a factor proportional to the square of the ratio of the final to the initial
value of $m_h$.

Given the relatively small size of their correction $\delta m_h$ one can
say that the particular CMSSM  points with $\Delta\!<\!100$
 and their predictions are rather stable against the presence of supersymmetric 
``new physics'' at the scale  $M = 8$ TeV. This finding
can be explained by the fact that these points generically have a light $\mu$  and
also light $m_{1/2}$, and thus the supersymmetric corrections $\delta m_h$, 
($\propto\mu^2\alpha_{j0}$, etc) are rather suppressed. The corrections $\delta m_h$ can 
increase  or decrease if one also includes effects  of Susy breaking associated to $\cO_i$, 
but these bring additional model dependence and extra parameters.

Let us now discuss about the CMSSM points in Figures~\ref{f1}, \ref{f2} 
with fine tuning $\Delta\!\geq \!200$, situated above
the upper continuous line. They can bring an increase of $m_h$ which can be
significant, of 10-30 GeV.
For example there are points which for $m_h$ near 100 GeV
can receive corrections of order 15-20 GeV, to  reach and comply with the LEP2 bound.
Interestingly, for $\cO_{1,2}$ the Higgs mass increase is such that total $m_h$ 
remains close to $\approx$122  GeV. Points that are largely fine-tuned and have 
a value for $m_h$ significantly below the LEP2 bound, are often receiving 
the largest corrections  $\delta m_h$. 
This opens the possibility that the phase space of the CMSSM be
increased and points which were  otherwise ruled out on grounds of extreme fine
tuning and/or LEP2 bound, can be  ``recovered'' and become phenomenologically viable.

Finally,  notice  that we kept all operators $\cO_j$ independent of each other. By doing 
so, one can single out the individual contributions of each operator (labelled by $\alpha_{j*}$),
which helps in model building, since not all operators are  present in a specific model. 
Also it is unlikely that ``new physics'' will bring up in the leading order, simultaneously, 
all these operators. What does all this mean for EW scale fine-tuning? 
The value of $\Delta$ for those points initially strongly fine-tuned 
can decrease by a factor equal to the square of the ratio  of the Higgs mass  after and 
before adding the correction $\delta m_h$ and this effect can be significant, as seen 
earlier for $\cL_1$ ($d=5$ case). A similar effect is expected for the case of 
d=6 operators. So one expects a change of $\Delta$ 
\bea
\Delta \ra \Delta \frac{m_h^2}{(m_h+\delta m_h)^2}
\eea
One  cannot obtain a more exact evaluation of $\Delta$ in our model-independent
approach, i.e. in the absence of the details of the new physics 
(quantum numbers of massive states) 
that generated the effective operators in the first instance. 
For further discussions on this see \cite{Antoniadis:2009rn}.

\subsubsection{Effective operators: removing redundant operators.}

The reader may notice that the discussion for the d=6 operators 
ignored a class of operators that could be present (and that are similar 
to $\cL_2$ of the d=5 case).  These operators involve 
extra derivatives and are
\medskip
\bea
&&
\!\!\!\!\!\!\!\!
\mathcal{O}_{9}=\frac{1}{M^{2}}\int d^{4}\theta \,\,
\mathcal{Z}_{9}\,H_{1}^{\dagger }\,
\overline{\nabla}^{2}\,e^{V_{1}}\,\nabla ^{2}\,H_{1}  
\qquad\quad\,\,\,\,
\mathcal{O}_{12} =\frac{1}{M^{2}}\int d^{4}\theta \,\,\mathcal{Z} 
_{12}\,H_{2}^{\dagger }\,e^{V_{2}}\,\nabla ^{\alpha 
}\,W_{\alpha }^{(2)}\,H_{2}  \label{der0} 
 \nonumber \\ 
&&\!\!\!\!\!\!\!\!
\mathcal{O}_{10}=\frac{1}{M^{2}}\int d^{4}\theta \,\,
\mathcal{Z}_{10}\,H_{2}^{\dagger }\,\overline{\nabla } 
^{2}\,e^{V_{2}}\,\nabla ^{2}\,H_{2} 
\qquad\quad
\mathcal{O}_{13} =\frac{1}{M^{2}}\int d^{4}\theta \,\,\mathcal{Z}
 _{13}\,H_{1}^{\dagger }\,e^{V_{1}}\,
 \,W_{\alpha }^{(1)}\,\nabla^\alpha\,H_{1}  
\nonumber \\ 
&&\!\!\!\!\!\!\!\!
\mathcal{O}_{11} =\frac{1}{M^{2}}\int d^{4}\theta \,\,\mathcal{Z}
_{11}\,H_{1}^{\dagger }\,e^{V_{1}}\,\nabla ^{\alpha 
}\,W_{\alpha }^{(1)}\,H_{1}  
\qquad
\mathcal{O}_{14}=\frac{1}{M^{2}}\int d^{4}\theta \,\,\mathcal{Z} 
 _{14}\,H_{2}^{\dagger }\,e^{V_{2}}\,
 \,W_{\alpha }^{(2)}\,\nabla^\alpha\,H_{2}
\nonumber\\
&& \!\!\!\!\!\!\!\!\!
\cO_{15}=\frac{1}{M^2} \int d^4\theta \textrm{Tr} e^V\,W^\alpha
e^{-V}\,D^2 (e^V\, W_\alpha\, e^{-V})
\,\,\,\, \label{der} 
\eea

\medskip\noindent
Here $\nabla_{\alpha }\,H_{i}=e^{-V_{i}}\,D_{\alpha }\,e^{V_{i}}
 H_i$ and  $W_\alpha^i$ is the field strength of $V_i$. $\cO_{15}$ does not
depend on Higgs explicitly, but could affect its scalar potential.
 To be general, in the above operators one should include 
spurion ($S$) dependence under any $\nabla_\alpha$  to account for 
supersymmetry breaking  effects associated to them. Such operators are easily 
generated when integrating out massive states in 4D models.  They can 
also be generated at one-loop by compactification, after integrating out 
towers of Kaluza-Klein states, in the presence of  localised (Yukawa) interactions 
\cite{Ghilencea:2004sq} (for example $\cO_9$, $\cO_{10}$)
or due to bulk (gauge) interactions \cite{GrootNibbelink:2005vi} ($\cO_{15}$).

Such operators are however redundant in the sense that they can be 
removed by non-linear  field redefinitions \cite{Antoniadis:2009rn,Antoniadis:2008es}.
To see  how this works, consider for example operator
$\cO_9$, without the gauge field dependence, and in the 
presence of an otherwise standard Lagrangian, with an 
arbitrary superpotential $W$:
\bea\label{rrr}
\cL= \int d^4\theta\,
\Big[\Phi^\dagger \,(1+\Box/M^2)\,\Phi+\chi^\dagger\chi\Big]
+\bigg\{\int d^2\theta \,\,
W[\Phi;\chi]+h.c.\bigg\}+\cO(1/M^3)
\eea
Here $\Phi^\dagger \Box\Phi$ comes from $\cO_9$ with the replacement $H_1\rightarrow \Phi$. 
Further, one replaces $\Phi^\dagger \Box\Phi\ra(-1/16)\overline D^2\Phi^\dagger D^2\Phi$.
This $\cL$ can be unfolded into a ``standard'' Lagrangian without extra
derivatives (see \cite{Antoniadis:2007xc}, also Appendix~B in the first paper in
 \cite{Antoniadis:2009rn}). 
To see how this works,  consider a change of basis to  $\Phi_{1,2}$:
$\Phi=s_1\Phi_1+s_2\Phi_2$ and $(1/m)\,\overline D^2\Phi^\dagger=r_1
\Phi_1+r_2\Phi_2$
where $s_{1,2}$, $r_{1,2}$ form an unitary matrix, so that the
eigenvalue problem is not changed; $m$ is a very small, non-zero
mass scale of the theory that can be taken to zero at the end of the
calculation. Since $\Phi$ and $\overline D^2\Phi$ are not independent, such
transformation must be accompanied by a Lagrangian constraint, which
must vanish in the limit $M\ra \infty$. This constraint has the form:
\bea
\delta\cL=\int d^2\theta\,\Big[(1/m)\,\overline
D^2(s_1\,\Phi_1+s_2\,\Phi_2)^\dagger-(r_1\Phi_1+r_2\,\Phi_2)
\Big]\big({m^2}/({4\,M})\big)\,\Phi_3
\eea
where $\Phi_3$ is a chiral superfield which plays the role of a Lagrange multiplier,
so that the total, equivalent Lagrangian in the new basis  is $\cL'\equiv \cL+\delta\cL$.
One then brings $\cL'$ to a diagonal form to find the result:
\bea\label{unfo}
\cL'\!=\!\!\int\!\! d^4\theta
\Big[\tilde \Phi_1^\dagger\tilde \Phi_1
\!-\!\tilde\Phi_2^\dagger\tilde  \Phi_2
\!-\!\tilde\Phi_3^\dagger\tilde  \Phi_3
\!+\!\chi^\dagger\chi\Big]
\!+\!\bigg\{\!\!\!\int\!\! d^2\theta \Big[W[\Phi(\tilde\Phi_{1,2});\chi]
\!-\! M\tilde\Phi_2\tilde\Phi_3\Big]\!+\!h.c.\!\!\bigg\}\!
+\!\cO\Big[\frac{1}{M^3}\Big]
\eea
where $\Phi=\tilde\Phi_2-\tilde\Phi_1$ and where  we took the
limit $m\ra 0$.
This Lagrangian is that of a second order theory, only polynomial in superfields, 
and classically equivalent to the initial one  (\ref{rrr}). Given the signs in the D-term, 
two massive superghosts ($\tilde\Phi_{2,3}$) are present, of mass near the effective 
cut-off, $\cO(M)$. Their presence is just  an effect of truncating the 
operators series expansion to $1/M^3$  terms\footnote{ 
Indeed, in a renormalisable or ghost-free theory with a massive state, when 
integrating out this state (by the eqs of motion) one finds in the low-energy 
effective action an infinite series of derivative terms, suppressed by powers of 
$M$. The ``truncated'' theory  will have a finite number of derivatives, and, 
as seen above, this will bring superghosts (or just ghosts in the non-Susy case), i.e. 
fields with negative kinetic term. This happens even though the 
theory with the infinite series is ghost-free, since the 
original theory was so.}. Also observe that the 
$\chi$ field was ``spectator'' throughout this analysis and did not
affect it; in fact the $\chi$-dependence can be replaced by an arbitrary
polynomial function. Further, since the superghost degrees of freedom  are 
massive, one can integrate them out, by  the equations of motion.  After a careful 
calculation and  consistent Taylor expansion,  
the result is (see appendix in  \cite{Antoniadis:2009rn})
\smallskip
\bea\label{hh}
\cL'\!=\!\int\! d^4\theta 
\Big[\tilde\Phi_1^\dagger\tilde\Phi_1\!-\!
\frac{1}{M^2}
 W^{\prime\,\dagger}[\tilde\Phi_1;\chi]
 W^\prime[\tilde\Phi_1;\chi]+\chi^\dagger\chi \Big]
\!+\!\bigg\{\int d^2\theta\,W[\tilde\Phi_1;\chi]
\!+\! h.c.\bigg\}\!+\!\cO\Big[\frac{1}{M^3}\Big]
\eea

\smallskip\noindent
where the derivatives of $W$ are taken wrt its first argument.
This Lagrangian contains only polynomial interactions (renormalisable
or not) and standard kinetic terms, and is equivalent to the original one, 
eq.(\ref{rrr}), but has no extra-derivative terms.

This result agrees with that obtained by using the equations of motion in the
derivative term in (\ref{rrr}), but this is not true in general, as we argue below.
Indeed, as shown in the appendix of  \cite{Antoniadis:2009rn}, the presence 
of a derivative term\footnote{Such term is part of  $\cL_2$ discussed
in the case of d=5 operators since
$\int d^2\theta \Phi\Box\Phi=-4 \int d^4\theta \,\Phi D^2\Phi$.}
 $\Phi\Box\Phi/M$ in the superpotential  of an otherwise arbitrary
Lagrangian leads, via the method shown above, to a result that is similar in order 
$\cO(1/M)$ to that obtained via the eqs of motion. However, in the $\cO(1/M^2)$ the result 
found by using ordinary eqs of motion is different and actually wrong. 
The discrepancy is due to the fact that Euler-Lagrange eqs are changed 
in the higher order theory, and this should be taken into account
when using eqs of motion to eliminate the higher-derivative operators. 
We find the method presented above more elegant and transparent.
  
Returning to the CMSSM with an additional $\cO_9$, the second term in (\ref{hh}) together
with the standard MSSM superpotential ($\sim \mu H_1.H_2$) can bring only wavefunction 
renormalization of the Higgs kinetic term or other effective operators 
polynomial in fields.  When Susy breaking is included, such terms also
bring soft terms and $\mu$-term redefinition.  A similar 
strategy applies to the rest of the operators listed in (\ref{der}).
Therefore all these operators are ``redundant'' in the sense that they can be eliminated, 
up to redefinitions of the fields, soft masses and $\mu$-term \cite{Antoniadis:2008es}. 
For this reason they were not considered in the phenomenological studies of 
the previous sections.

\section{Conclusions}

A quantitative test of supersymmetry as a solution to the hierarchy problem is
the fine-tuning $\Delta$  of the electroweak scale with respect to variations 
of the UV parameters of the model, after including the quantum corrections
 and experimental and theoretical constraints. While the largest value of 
$\Delta$ for which Susy is still a solution to this 
problem is somewhat subjective, it is very natural to ask that  
$\Delta$ be minimal in a given model. This point of view is also
supported in part by Bayesian approaches to data fits, 
in which $\Delta$ is automatically included in the {\it effective} prior
expression as an extra  $1/\Delta$ factor, so that the method brings 
naturally a fine-tuning 
penalty for points of large $\Delta$. This underlines the   physical meaning of 
$\Delta$ in general and the need of minimizing the overall fine-tuning as done in this work,  
to select the points of physical relevance.

We applied this idea of 
minimizing $\Delta$ to the CMSSM at two-loop leading log, and investigated its
results for the value of the radiatively corrected value of the lightest Higgs mass, $m_h$.
Remarkably, although $\Delta$ depends $\approx$ exponentially on $m_h$, there 
does exist a minimum at a very acceptable value ($\Delta=8.8$) corresponding to a mass
 $m_h=114\pm 2$ GeV.  The very existence of such a minimum situated at the 
intersection of two {\it exponential} dependences on both sides of this value for $m_h$, and induced 
by quantum corrections, cannot be stressed enough. The exponential growth of $\Delta$ for $m_h$ 
above this value is largely due to QCD quantum effects which overcome the ``good'' 
Yukawa loop effects needed to induce radiative EW symmetry breaking. Thus QCD ``does not like''
a larger $m_h$ unless one is prepared to pay the associated (high) fine-tuning cost. Imposing  
consistency with the WMAP bound, 
the above value changes mildly to $\Delta =15$ leading to $m_h=114.7\pm 2$ GeV, 
while saturating this bound within 3$\sigma$ leads to $\Delta=17.8$ 
with $m_h=115.9\pm 2$ GeV (the quoted theoretical uncertainty ($\pm 2$ GeV) can actually 
be larger, up to $3$ GeV).

It is indeed remarkable that constraints from short distance physics (EW precision data) and
from large distance physics (dark matter) can be consistent with each other so accurately,
and together can help one to make physical predictions.
We also checked that using a different definition for $\Delta$, such
as $\Delta^\prime=\sqrt{\sum_i\Delta_i^2}$
does not change this result for $m_h$, since $\min\Delta^\prime$ is found at similar values
for $m_h$ and its plot as a function of $m_h$ is very similar.
Finally, let us note that points with $\Delta<100$ are currently  being 
tested by  the LHC 7 TeV run and by dark matter experiments (CDMS, Xenon, etc)
and their impact was briefly discussed.

We further discussed how the results for $\Delta$ and $m_h$ change 
under the addition of ``new physics'' beyond   the MSSM Higgs sector
and also whether one could have a larger $m_h$   with a low $\Delta$. 
This is relevant for MSSM since for $m_h>121$ GeV, $\Delta$ is already $\Delta>100$, 
and becomes 1000 for $m_h=126$ GeV!  
The ``new physics''  could reduce $\Delta$ to acceptable values, even at larger $m_h$.
It could be represented by additional Higgs doublets, massive 
gauge singlets or massive U(1)$^\prime$, etc, and its effects can be generally parametrized 
by series of effective operators, suppressed by the scale of massive states that generated 
them. We considered both d=5 and d=6 operators. In the presence alone of the former, one can 
achieve a low $\Delta<10$ even for $m_h$ as large as $\approx 130$ GeV. 
Such a $d=5$ effective operator can be generated by integrating out a 
massive gauge singlet beyond MSSM, whose effect is to increase the quartic higgs coupling
and reduce the fine tuning for similar $m_h$. This case is nothing but 
the decoupling limit of the generalised NMSSM which contains a supersymmetric
mass term for the gauge singlet (in ordinary NMSSM, the fine tuning reduction is 
smaller). Other ways to generate such operator can however exist.

Regarding the operators of dimension d=6, we considered their individual, supersymmetric 
corrections to the CMSSM 2-loop  leading log value of $m_h$, for an expansion parameter  
$\tilde m/M<1/4$ with $\tilde m$ any scale of the model $(m_0,m_{1/2},\mu,v)$. 
Their effects were treated as a perturbation of the CMSSM ``background'' points 
that respected all experimental constraints. It was shown that CMSSM points with $\Delta<100$ 
having $m_h<121$ GeV ($\Delta<200$, $m_h<122$ GeV) receive supersymmetric corrections of up 
to 4 (6) GeV from individual operators, respectively. Therefore, the points in the 
CMSSM of lowest fine-tuning receive rather modest (supersymmetric) corrections from
 individual operators, and are therefore rather stable under "new physics" at a scale
 considered here of 8 TeV. Applied to the point of minimal fine tuning ($114\pm 2$GeV), 
the correction mentioned of 4(6) GeV from individual d=6 operators brings $m_h$  up to 
a value of $118(120)\pm 2$ GeV. Including a 3$\sigma$ consistency with WMAP changes 
this result mildly  up to $119.9 (121.9)\pm 2$ GeV. Finally, an increase (decrease)
 of the scale $M$ by 1 TeV brings a decrease (increase) of the correction by about 1 GeV.
 A next step in this study is to impose
 dark matter constraints  on the CMSSM with d=6 operators, since by supersymmetry
 these extend the neutralino sector too. The obtained bounds on $M$ can then be
 used to re-evaluate   the quoted  correction to $m_h$ in the presence of these operators.

\section*{Acknowledgments}

The work of SC was supported by the UK Science and Technology 
Facilities Council under contract PPA/S/S/2006/04503.
This work was supported in part by a grant of the Romanian National Authority 
for Scientific Research, CNCS – UEFISCDI, project number 
PN-II-ID-PCE-2011-3-0607.

\end{document}